\documentclass[11pt]{article}

\usepackage{graphicx}
\usepackage{mathrsfs}
\usepackage{latexsym}
\usepackage{amsmath}
\usepackage{amssymb}
\usepackage{color}
\usepackage{subfig}
\usepackage{array}
\usepackage{multirow}

\usepackage[nohead,letterpaper]{geometry}
\usepackage{vmargin}
\setpapersize{USletter}
\setmargrb{2cm}{1cm}{2cm}{1.75cm}

\usepackage{xcolor}
\definecolor{plum}{rgb}{0.36078, 0.20784, 0.4}
\definecolor{chameleon}{rgb}{0.30588, 0.60392, 0.023529}
\definecolor{cornflower}{rgb}{0.12549, 0.29020, 0.52941}
\definecolor{scarlet}{rgb}{0.8, 0, 0}
\definecolor{brick}{rgb}{0.64314, 0, 0}

\newcommand{\email}[1]{\href{mailto:#1}{\tt \textcolor{cornflower}{#1}}}

\newcommand{\ba}{\begin{eqnarray}}
\newcommand{\ea}{\end{eqnarray}}
\newcommand{\be}{\begin{equation}}
\newcommand{\ee}{\end{equation}}
\newcommand{\bd}{\begin{displaymath}}
\newcommand{\ed}{\end{displaymath}}

\usepackage[%
  pdftitle={Generalization of Deformations of Lifshitz holography into higher dimensions and Lovelock gravity},%
  pdfauthor={Miok Park, Robert Mann},%
  pdfsubject={Generalization of Deformations of Lifshitz holography into higher dimensions},%
  pdfkeywords={Lifshitz, Holographic Renormalization, Boundary Terms, Surface Terms, Variational
    Principle},%
  colorlinks=true,linkcolor=cornflower,citecolor=brick,urlcolor=chameleon%
  ]{hyperref}

\numberwithin{equation}{section}

\usepackage[bf,medium,]{titlesec}

\titleformat{\section}
{\large\bfseries}{\thesection.}{.5em}{\large\bfseries}

\titleformat{\subsection}
{\bfseries}{\thesubsection}{.5em}{\bfseries}

\begin{document}


\thispagestyle{empty}

\begin{flushright}
\today
\end{flushright}
~\vspace{2cm}\\

\begin{center}
{\bf \Large Deformations of Lifshitz Holography in $(n+1)$-dimensions }
\end{center}

\vspace{.5cm}

\begin{center}
Miok Park $^{a}$ and Robert B. Mann $^{a, b}$

\vspace{1cm}{\small {\textit{$^{a}$Department of Physics,\\ University of
Waterloo,\\ Waterloo, Ontario N2L 3G1,\\ Canada}}}\\
\vspace{2mm} {\small {\textit{$^{b}$Perimeter Institute for Theoretical Physics,\\ 31 Caroline Street North,\\ Waterloo,
Ontario N2L 2Y5,\\ Canada}}}\\
\vspace*{0.5cm}
\email{m7park@uwaterloo.ca}\,,
\email{rbmann@uwaterloo.ca}
\end{center}
\vspace{0.5cm}

\begin{center}
{\bf Abstract}
\end{center}

 We investigate deformations of Lifshitz holography in $(n+1)$ dimensional spacetime.  After discussing the situation for
general Lifshitz scaling symmetry parameter $z$, we consider $z=n-1$ and the associated marginally relevant operators. These operators are dynamically generated by a momentum scale $\Lambda \sim 0$ and correspond to   slightly deformed Lifshitz spacetimes via a holographic picture. We obtain  renormalization group flow at   finite temperature from UV Lifshitz to IR AdS, and
evaluate how  physical quantities such as the free energy density and the energy density depend on $\log(\Lambda^z/T)$ in the quantum critical regime as $\Lambda^z/T \rightarrow 0$.

\newpage

\setcounter{page}{1}

\tableofcontents


\newpage

\section{Introduction}
\label{sec:Intro}

One of the most innovative ideas in theoretical physics in recent years is the anti-de Sitter spaceetime/Conformal Field Theory (AdS/CFT) correspondence\cite{Maldacena:1998}, which posits an isomorphism between symmetries of a spacetime and those of matter fields -- for instance, the conformal group $SO(n,2)$ of a $n$-dimensional CFT aries as the group of isometries of $AdS_{n+1}$. This idea has been extended to a duality between gauge theories and gravitational theories in one larger dimension,
yielding a new way to understand physics. The most exciting aspect of this picture is to provide a technique for obtaining a weakly  coupled and calculable dual description of strongly coupled matter fields  in terms of gravity.

Several years prior to the conception of the AdS/CFT correspondence in high energy physics,  investigations on phase transitions of  modern materials indicated that a precarious point exists between two stable phases of matter such as superconductors and ferroelectrics or ferromagnets in which the temperature of a system has been driven to absolute zero by the application of some
external parameter such as pressure or an applied magnetic  field.  Unlike classical critical points, where   critical fluctuations are limited to a narrow region around the phase transition, at such  'quantum critical points', the critical fluctuations are quantum mechanical in nature and exhibit a generalized scale invariance in both time and space. Understanding this puzzling behavior has become a major research effort in condensed matter physics. Such systems exhibit universally distinct characteristics upon fanning out to finite temperatures, and so
the effect of quantum criticality is felt without ever reaching absolute zero.   A recent attempt to understand  quantum
critical theory involves extending the AdS/CFT correspondence \cite{Son:2008}, \cite{Gubser:2008} to
condensed matter systems, yielding a considerably broader range of scope for gauge/gravity duality.

The signature scaling property underlying  quantum critical theory in $2+1$ dimensions is
\begin{equation}\label{QCTscale}
t \rightarrow \lambda^z t, \; \; \; \; \vec{x} \rightarrow \lambda \vec{x}
\end{equation}
where $z$ is the dynamical critical exponent;  $z=1$ corresponds to conformal invariance, whereas $z\neq 1$ implies an  anisotropic scaling invariance. Recently a form of gauge-gravity duality was proposed for  $z\neq 1$
 \cite{Kachru:2008}, in which the geometrical dual is obtained from the (asymptotic) metric
\begin{equation}\label{KLMmetric}
ds^2 = l^2 \bigg(- \frac{dt^2}{r^{2z}} + \frac{dr^2}{r^2} + \frac{dx^2 + dy^2}{r^2} \bigg)
\end{equation}
which is called Lifshitz spacetime and obviously satisfies
\begin{equation}\label{Lifscale}
t \rightarrow \lambda^z t, \; \; \; \; r \rightarrow \lambda r,\; \; \; \; \vec{x} \rightarrow \lambda \vec{x}.
\end{equation}
When $z=1$ the metric (\ref{KLMmetric}) is that of (asymptotic) AdS spacetime and (\ref{QCTscale}) recovers the conformal symmetry of the CFT. When $z=2$, (\ref{QCTscale}) restores the scaling symmetry  of quantum critical theories. The more general anisotropic scaling symmetry, $z \neq 1$, submerged in the gravity theory and the field theory, is the foundation for a Lifshitz spacetime/Quantum Critical Theory (Lifshitz/QCT) correspondence.

One issue associated with this approach  is how to obtain non-trivial spacetimes that asymptote  to the  anisotropic metric (\ref{KLMmetric}). Two approaches have been  considered to this end. First, it is obvious from the Einstein equations that an anisotropic energy-momentum tensor could support  an anisotropic geometry; for example   a massive vector field with the appropriate asymptotic behaviour can suffice.  An alternate approach involves   adding higher curvature terms into the Einstein action \cite{Dehghani:2010}; by appropriately tuning the different gravitational constants, metrics asymptotic to (\ref{KLMmetric}) can be obtained. In this paper, we follow the first approach, investigating the Einstein action coupled to a massive vector field in $(n+1)$ dimensions.

Related to Lifshitz field theory, an interesting feature attracting much recent attention is associated with renormalization group flow.  In $z=2$ Lifshitz theory, the action is
\begin{equation}
S_{Lif} = \frac{1}{2} \int d \tau d^2 x \bigg( (\partial_{\tau} \phi)^2 - \kappa (\nabla^2 \phi)^2 \bigg),
\end{equation}
which  has an anisotropic scaling invariance. It is known that action describes  strongly correlated electron systems;   its fixed points seem to flow to a non-Abelian gauge theory by perturbing the action with a term $-(\nabla \phi)^2$. From the perspective of holographic duality, we expect holographic renormalization flow from a UV-Lifshitz fixed point to an AdS fixed point under the relevant perturbation, a result obtained numerically in {\cite{Kachru:2008}}.  To obtain a metric asymptotic to (\ref{KLMmetric}), a Proca field is necessary {\cite{Kachru:2008}}; its essential physics for $z = 2$ in (3+1) dimensions is that of  a marginally relevant operator in the quantum critical theory, which induces a flow from the $z = 2$ theory to a relativistic $z = 1$ infrared fixed point.   Advancing this study further, the effect of such
marginally relevant operators  at finite temperature was recently explored, with the renormalization flow for UV Lifshitz to IR AdS described and the physics explored in the quantum critical regime {\cite{Cheng:2010}}.

In this paper, we consider $(n+1)$-dimensional Lifshitz spacetime and $((n-1)+1)$-dimensional Quantum Critical Theory(QCT), and study their holographic duality.  While QCT is well described in a $2+1$ dimensional context, more general theories of physics including the standard model and gravity are implemented in a higher-dimensional context.  The success of the AdS/CFT correspondence therefore provides motivation to understand the extent to which the broader notions of Lifshitz/QCT duality are applicable in higher dimensions, and  what different behaviour emerges. Motivated by these interests, we especially focus on the marginally relevant operators in the QCT extended to higher dimensions, with the goal of understanding their behaviour from the perspective of holographic duality, where these operators correspond to the deformed Lifshitz spacetime solutions.

Previous work in this subject has concentrated on the (2+1)-dimensional case \cite{Cheng:2010}. Here we demonstrate that
 renormalization group flow  from Lifshitz spacetime in the UV to AdS spacetime in the IR generalizes to any dimensionality
in the marginally relevant case  yielding deformations of the pure Lifshitz spacetime. From a thermodynamic
 perspective, we find that  physical quantities such as the ratios $s/T$  (entropy density to temperature),
 ${\mathcal F}/Ts$  (free energy density to $Ts$), and ${\mathcal E}/Ts$ (energy density over $Ts$) exhibit   progressively
weaker dependence on temperature at sub-leading order in $\log(\Lambda^z/T)$ as dimensionality increases.  We also find
that  the maximal flux of the vector field near the horizon grows linearly with increasing dimension.

In section 2, the action, equations of   motion, and basic setup are introduced, along with an ansatz for which all constants are fine-tuned and normalized for both Lifshitz and AdS spacetime.   In section 3, asymptotic solutions consistent with a marginally relevant operator are derived by bringing in a dynamically generated momentum scale $\Lambda$ (assumed very small), which deforms Lifshitz spacetime in the high energy regime.  In section 4, we carry out
holographic renormalization,  rendering the action finite by constructing  proper counterterms.   In section 5, we numerically match our asymptotic near-Lifshitz solutions with black hole solutions near the horizon.   We then describe the renormalization group flow, and compute physical quantities such as the entropy density $s$, the free energy density ${\mathcal{F}}$, and the energy density $\mathcal{E}$ for $n=3,4,5,6,7,$ and $8$.

\section{Einstein Gravity with a Massive Vector fields in $(n+1)$ dimensional spacetime}
\label{HD expn}

The  action for gravity in $(n+1)$-dimensional spacetime coupled to a massive vector field is described by
\begin{equation}
S = \int d^{n+1} x \sqrt{-g} \bigg( \frac{1}{2 {\kappa_{n+1}}^2} [R + 2 \tilde{\Lambda}] - \frac{1}{{g_v}^2} \bigg[ \frac{1}{4} H^2 +
\frac{\gamma}{2} B^2 \bigg] \bigg)
\end{equation}
where $\kappa_{n+1} = \sqrt{8 \pi G_{n+1}}$ in which $G_{n+1}$ is the $(n+1)$ dimensional gravitational constant, and $H=dB$ and
$g_v$ is the (n+1) dimensional coupling constant of the vector field. The equations of the motion are
\begin{equation}
\frac{1}{{\kappa_{n+1}}^2} \bigg( R_{\mu \nu} - \frac{1}{2} g_{\mu \nu} R - \tilde{\Lambda} g_{\mu \nu} \bigg) = \frac{1}{{g_v}^2} \bigg( H_{\mu \rho} H_{\nu}^{\; \rho} - \frac{1}{4} g_{\mu \nu} H^2 \bigg) + \frac{\gamma}{{g_v}^2} \bigg( B_{\mu}B_{\nu} - \frac{1}{2} g_{\mu \nu} B^2 \bigg),
\end{equation}
and
\begin{equation}
\nabla_{\mu} H^{\mu \nu} - \gamma B^{\nu} = 0
\end{equation}
where $\gamma$ is the squared mass of the vector field. For the action to yield solutions asymptotic to those having the scaling symmetry (\ref{Lifscale}), we require   the spacetime metric
\begin{equation}\label{ndimmet}
ds^2 = l^2 \bigg( - \frac{dt^2}{r^{2z}} + \frac{dr^2}{r^2} + \frac{dx^2 + dy^2 + \cdots}{r^2} \bigg)
\end{equation}
to be a solution to the field equations,  where $z$ is arbitrary. Note that in these coordinates   $r \rightarrow 0$ corresponds to the boundary of the spacetime.

The vector potential yielding a stress-energy supporting this metric is given by \begin{equation}
B = \frac{g_v l}{\kappa_{n+1}} \frac{q}{r^z} dt.
\end{equation}
These ansatz and   boundary conditions fine-tune the cosmological constant to be
\begin{equation}\label{Lamfine}
\tilde{\Lambda} = \frac{(z-1)^2 + n(z-2)+n^2}{2 l^2},
\end{equation}
and the  squared mass and the charge of the vector field to be
\begin{equation}\label{gamfine}
\gamma = \frac{(n-1)z}{l^2}, \; \; \; \; \; \; \; \; \; \; \; \; \; q^2 = \frac{z-1}{z}.
\end{equation}
Regardless of the dimensionality of the spacetime, setting $z=1$ in (\ref{ndimmet}) yields $AdS_{n+1}$ solution
\begin{equation}\label{AdSmet}
ds_{AdS}^2 = a \; l^2 \bigg( - \frac{dt^2}{r^2} + \frac{dr^2}{r^2} + \frac{dx^2 + dy^2 + \cdots}{r^2} \bigg)
\end{equation}
where the vector potential vanishes.
As the cosmological constant has been already fixed due to the Lifshitz boundary condition
we introduce a scaling constant, $a$, into the AdS metric and adjust its value to be
\begin{equation}\label{ascale}
a = \frac{n(n-1)}{(z-1)^2 + n(z-2) + n^2}.
\end{equation}
Once we fix the cosmological constant (\ref{Lamfine}) with space dimension  $n$ and dynamical critical exponent $z$, then those values determine the scaling constant for the AdS spacetime metric.

In order to describe the renormalization group flow which involves   breaking the anisotropy of the spacetime by running
from the UV Lifshitz to the IR AdS, we employ the ansatz
\begin{eqnarray}
ds^2 &=& l^2 \bigg(-f(r)dt^2 + \frac{dr^2}{r^2} +p(r) (dx^2 + dy^2 + \cdots) \bigg), \label{metanz}\\
B &=& \frac{g_v  l}{\kappa_{n+1}} h(r) dt  \label{Banz}
\end{eqnarray}
so for the Lifshitz spacetime
\begin{equation}
\textrm{Lifshitz : } f = \frac{1}{r^{2z}} \; \; \; p = \frac{1}{r^2}, \; \; \; h= \frac{\sqrt{z-1}}{\sqrt{z}} \frac{1}{r^z},
\end{equation}
whereas for the $AdS_{n+1}$ spacetime
\begin{equation}
\textrm{AdS : } f = p = \frac{n}{(3n-4)} r^{-2 \sqrt{\frac{3n-4}{n}}}, \; \; \; h= 0.
\end{equation}
With (\ref{metanz}) and (\ref{Banz}), the equations of motion yield  three independent non-linear ODEs for $\{f(r),p(r),h(r)\}$
\begin{eqnarray} \label{eqfph}
&&2 \chi + \frac{z(4n-6) h(r)^2}{f(r)} - \frac{r f'(r)}{f(r)} + \frac{r^2 f'(r)^2}{2 f(r)^2} - \frac{(3n -5) r^2 f'(r) p'(r)}{2
f(r) p(r)} - \frac{(n-2)^2 r^2 p'(r)^2}{2 p(r)^2} - \frac{r^2 f''(r)}{f(r)} = 0, \nonumber\\
&&-\frac{2 z h(r)^2}{f(r)} - \frac{r p'(r)}{p(r)} + \frac{r^2 f'(r) p'(r)}{2 f(r) p(r)} + \frac{r^2 p'(r)^2}{2 p(r)^2} - \frac{r^2
p''(r)}{p(r)} = 0, \nonumber\\
&& \chi + \frac{(n-1)z h(r)^2}{f(r)} - \frac{r^2 h'(r)^2}{f(r)} - \frac{(n-1) r^2 f'(r) p'(r)}{2 f(r) p(r)} - \frac{(n-2) (n-1) r^2
p'(r)^2}{4 p(r)^2} = 0,
\end{eqnarray}
where $\chi  = (n-1)^2 + (n-2) z + z^2$. We shall rewrite the equations of the motion with the new variables
\begin{equation}\label{newvars}
p(r) = e^{\int^{r} \frac{q(s)}{s} ds}, \; \; \; \; f(r) = e^{\int^{r} \frac{m(s)}{s} ds}, \; \; \; \; h(r) = k(r) \sqrt{f(r)}.
\end{equation}
These variables  have the added benefit of turning the second order differential
equations into  first order and  postponing the determination of rescaling ambiguities on $f$, $p$, and $h$.

For further simplification, we introduce a new variable
\begin{equation}\label{newx}
x(r) = \bigg( 4 \chi + 4(n-1)z k(r)^2- 2(n-1)m(r)q(r) - (n-2) (n-1)q(r)^2 \bigg)^{\frac{1}{2}}.
\end{equation}
Putting (\ref{newvars}) and (\ref{newx}) into (\ref{eqfph}) gives
\begin{eqnarray}
r x'(r) &=& -2 (n-1) z k(r) - \frac{(n-1)}{2} q(r)x(r), \nonumber\\
r q'(r) &=& \frac{\chi}{(n-1)} - z k(r)^2 - \frac{n}{4} q(r)^2 - \frac{1}{4 (n-1)} x(r)^2, \nonumber\\
r k'(r) &=& - \frac{\chi }{(n-1)} \frac{k(r)}{q(r)} - \frac{z k(r)^3}{q(r)} + \frac{(n-2)}{4} k(r)q(r) - \frac{x(r)}{2} +
\frac{1}{4 (n-1)} \frac{k(r) x(r)^2}{q(r)} \label{eqkqx}.
\end{eqnarray}
In terms of the new variables $k(r)$, $q(r)$, and $x(r)$, Lifshitz spacetime is described by
\begin{equation}
\textrm{Lifshitz : } q = -2 \; \; \; x = 2 \sqrt{z-1} \sqrt{z}, \; \; \; k= \frac{\sqrt{z-1}}{\sqrt{z}},
\end{equation}
whereas for AdS spacetime
\begin{equation}
\textrm{AdS : } q= -2 \sqrt{\frac{3n-4}{n}}, \; \; \;x=0, \; \; \; k= 0.
\end{equation}
So far we have worked with a general value of $z$ in $(n+1)$ dimensions. We are interested in studying the effects of marginal operators, which have  scaling dimension $z+n-1$ in Lifshitz spacetime, because of the different scaling of the time coordinate. While it has been shown that  the linearized equation of motion for the scalar part of constant perturbations in a Lifshitz background  in (3+1) dimensionals {\cite{Ross:2009ar}} (and the gravitational field has solutions that are marginal for  general $z$),  the vector field only admits a single degenerate solution at the special value of $z=2$, where the vector operator also becomes marginal. Applying this analysis to $(n+1)$ dimensions {\cite{BravinerGregoryRoss:2011aug}}, the condition for having a single degenerate  solution for the vector field is $z=n-1$, and the operators with this value are considered to be marginal.  Henceforth we deal with the case satisfying $z = n-1$.

\section{Asymptotic Behaviour}
\label{HD expn}

We consider the spacetime slightly thermally heated and so slightly deformed from the pure Lifshitz case, restricting our considerations to  $z=n-1$ for which the massive vector field becomes marginal. Under these assumptions, the general form of the solutions near the boundary $r \rightarrow 0$ is
\begin{align}
k(r)&=\frac{\sqrt{z-1}}{\sqrt{z}} \bigg(1+ \frac{1}{(z-1)^2 \log(r \Lambda)} + \frac{(z-1)(-3z + 2(z-1)^3 \lambda) +2 (1-3z)
log(- \log(r \Lambda))}{2 z(z-1)^4 \log^2 (r \Lambda)} + \cdots \bigg)\nonumber\\
&+ (r \Lambda)^{2z} \log^2(r \Lambda) \bigg( \beta \bigg( 1 + \frac{2(3z-1) \log(- \log(r \Lambda))}{z (z-1)^2 \log(r \Lambda)} + \cdots
\bigg) + \alpha \bigg( \frac{1}{ \log(r \Lambda)} + \nonumber\\
&\frac{(2z^2-4z+1) -2(z-1)^4 (2z-1)\lambda + 2 (6z^2 -5z+1) \log(- \log(r \Lambda)) }{2 z (z-1)^2 (2z-1) \log^(r \Lambda)} + \cdots\bigg) \bigg), \label{AsympSolk}\\
q(r)&=-2 \bigg(1-\frac{1}{(z-1) \log(r \Lambda)} - \frac{z+2(z-1)^4 \lambda -2(3z-1) \log(- \log(r \Lambda))}{2 z (z-1)^3 \log^2(r
\Lambda)} + \cdots \bigg) \nonumber\\
&-\frac{2 \sqrt{z-1} \sqrt{z}}{2z-1} (r \Lambda)^{2z} \log^2(r \Lambda) \bigg( \beta \bigg( 1+\frac{-z(4z^2 -7z
+2)+2(2z-1)(3z-1) \log(- \log(r \Lambda))}{z(z-1)^2(2z-1) \log(r \Lambda)} + \cdots \bigg) \nonumber\\
&+ \alpha \bigg(\frac{1}{ \log(r \Lambda)} - \frac{(2z^2-4z+1) +2 (z-1)^4 \lambda - 2 (3z-1) \log(- \log(r \Lambda))}{2z(z-1)^2 \log^2 (r
\Lambda)} + \cdots \bigg) \bigg), \label{AsympSolq}\\
x(r)&=2 \sqrt{z-1} \sqrt{z} \bigg(1 + \frac{z}{(z-1)^2 \log(r \Lambda)} + \frac{(z-1)^4 \lambda + (1-3z) \log(- \log(r
\Lambda))}{(z-1)^4 \log^2 (r \Lambda)} + \cdots \bigg) \nonumber\\
&-\frac{2z^2}{2z-1} (r \Lambda)^{2z} \log^2(r \Lambda) \bigg( \beta \bigg(1 + \frac{-z(4z^2-5z+1) + 2(6z^2 - 5z +1) \log(- \log(r
\Lambda))}{z(z-1)^2(2z-1) \log(r \Lambda)} + \cdots \bigg) \nonumber\\
&+ \alpha \bigg(\frac{1}{ \log(r \Lambda)} -\frac{(2z-1)^2 +2(z-1)^4 \lambda - 2(3z-1) \log(- \log(r \Lambda))}{2z(z-1)^2 \log^2 (r\Lambda)} + \cdots \bigg) \bigg), \label{AsympSolx}
\end{align}
where $\Lambda$ is a momentum scale, generating a marginally relevant mode, whereas $\Lambda^z$ is an energy scale with $n=z+1$   the spatial dimension.  As $\Lambda\to 0$ the solution recovers the pure Lifshitz
spacetime. The parameters $\alpha$ and $\beta$ describe other modes of the solution, and  $\lambda$ is nothing but a 'gauge choice' \cite{Cheng:2010}. In other words $\lambda$ is related to defining the scale $\Lambda$, and the solution $\{ k, q, x \}$ transforms as
\begin{equation}
F(\Lambda r; \alpha, \beta ; \lambda) = F(e^{\lambda'/z} \Lambda r; e^{-2 \lambda'} (\alpha - \lambda' \beta), e^{-2 \lambda'} \beta; \lambda + \lambda')
\end{equation}
where $F$ stands for the $k,q$ and $x$ functions. This is easily verified by noting that the solutions $k,q$ and $x$ with $\lambda=0$ can be obtained by setting $\lambda' = - \lambda$, and replacing $\Lambda r$, $\alpha$, and $\beta$ with $e^{- \lambda/z} \Lambda r$, $e^{2 \lambda} (\alpha + \lambda \beta)$, and $e^{2 \lambda} \beta$ respectively, and then re-expanding the solutions under the assumption $|\log(\Lambda r)| \gg |\lambda|$. Here we fix this ambiguity by setting $\lambda = 0$.

As we are interested in the high energy regime, we  expand by introducing an arbitrary scale $\mu$ and write
\begin{equation}
\log(r \Lambda) = \log(r \mu) - \log \frac{\mu}{\Lambda}.
\end{equation}
In the high energy regime where $\mu \gg \Lambda$ we have
\begin{equation} {\label{Expsn}}
\bigg | \frac{1}{ \log \frac{\mu}{\Lambda}} \bigg| , \; \; \; \;  \bigg | \frac{ \log(r \mu)}{ \log \frac{\mu}{\Lambda}} \bigg| \leqslant 1 .
\end{equation}

Upon expansion, equations (\ref{AsympSolk}) $\sim$ (\ref{AsympSolx}) become
\begin{align}
&k(r)=\frac{\sqrt{z-1}}{\sqrt{z}} \bigg(1 + \frac{1}{(z-1)^2 \log(\frac{\mu}{\Lambda})} - \frac{3z(z-1) + 2z(z-1)^2 \log(r \mu) +
 2 (3z-1) \log(- \log(\frac{\mu}{\Lambda}))}{2z(z-1)^4 \log^2(\frac{\mu}{\Lambda})} + \cdots \bigg), \label{HgExpk}\\
&q(r)=-2 \bigg(1-\frac{1}{(z-1) \log(\frac{\mu}{\Lambda})} + \frac{-z+2z(z-1)^2 \log(r \mu) +
 2(3z-1) \log(- \log(\frac{\mu}{\Lambda}))}{2z(z-1)^3 \log^2(\frac{\mu}{\Lambda})} + \cdots \bigg), \label{HgExpq}\\
&x(r)=2 \sqrt{z-1}\sqrt{z} \bigg(1+ \frac{z}{(z-1)^2 \log(\frac{\mu}{\Lambda})} -\frac{z(z-1)^2 \log(r \mu) + (3z-1)
log(- \log(\frac{\mu}{\Lambda}))}{(z-1)^4 \log^2(\frac{\mu}{\Lambda})} + \cdots \bigg). \label{HgExpx}
\end{align}
Using these solutions for $k(r), q(r)$ and $x(r)$, we employ the change of variables (\ref{newvars}) and (\ref{newx}) in reverse to obtain the original form  of the solutions
\begin{align}
f(\rho)&=\frac{F^2_{0}}{(r \Lambda)^{2z} (- \log(r \Lambda))^{\frac{2z}{z-1}}} \bigg(1 - \frac{(7z-4)+2(3z-1) \log(- \log(r
\Lambda))}{(z-1)^3 \log(r \Lambda)} - \frac{(23z^4-142z^3+152z^2-57z+6)}{4z(z-1)^6 \log^2 (r \Lambda)}\nonumber\\
&+\frac{(3z-1)^2 (5z-2) \log(- \log(r \Lambda)) + (3z-1)^3 \log^2(- \log(r \Lambda))}{z(z-1)^6 \log^2 (r \Lambda)} + \cdots \bigg) \label{AsympSolf}\\
p(\rho)&=\frac{P^2_{0}(- \log(r \Lambda))^{\frac{2}{z-1}}}{(r \Lambda)^2} \bigg(1 + \frac{(5z-2) + 2(3z-1) \log(- \log(r
\Lambda))}{z(z-1)^3 \log(r \Lambda)} + \frac{(31z^4-64z^3+106z^2-69z+14)}{4 z^2 (z-1)^6 \log^2(r \Lambda) } \nonumber\\
&+\frac{(3z^3+26z^2-21z+4) \log(- \log(r \Lambda))+(3z-1)^2 (z-3) \log^2(- \log(r \Lambda)) }{z^2 (z-1)^6 \log^2(r \Lambda)} + \cdots\bigg) \label{AsympSolp}
\end{align}
where $F_{0}$ and $P_{0}$ are constants.  Furthermore, in  the high energy regime the same expansion for eq. (\ref{AsympSolf}) --(\ref{AsympSolp}) yields
\begin{align}
&f(\rho) = \frac{1}{r^{2z}} \bigg(1 + \frac{7z-4+2z(z-1)^2\log(r \mu)+2(3z-1) \log(\log(\frac{\mu}{\Lambda}))}{(z-1)^3
log(\frac{\mu}{\Lambda}) } + \cdots \bigg), \label{hgexpf}\\
&p(\rho) = \frac{1}{r^2} \bigg(1-\frac{5z-2+2z(z-1)^2 \log(r \mu)+2(3z-1) \log(\log(\frac{\mu}{\Lambda}))}{z(z-1)^3
\log(\frac{\mu}{\Lambda})} + \cdots \bigg) \label{hgexpp}
\end{align}
upon  rescaling the $t$ and $x$ coordinates to
\begin{align} \label{scalingtx}
t \rightarrow \frac{\bigg( \Lambda \; \log^{\frac{1}{z-1}}(\frac{\mu}{\Lambda}) \bigg)^z}{F_{0}} t, \; \; \; \; x \rightarrow \bigg(\frac{\Lambda}{ \log^{z-1}(\frac{\mu}{\Lambda})} \bigg)^2 \frac{1}{P_{0}} x .
\end{align}

\section{Holographic Renormalization}
\label{HD expn}

In this section we investigate  thermodynamic quantities such as free energy density or energy density at an asymptotic boundary of the deformed Lifshitz spacetime. We begin with the definition of the free energy density
\begin{equation}
F = - T \log \mathcal{Z} = T S_{\epsilon} (g_{*})
\end{equation}
where $S_{\epsilon}$ and $g_{*}$ are respectively the Euclidean action and the metric,  and $\mathcal{Z}$ is the partition function.
Upon carrying out a variation of the on-shell action, boundary terms arise, and to cancel these out a Gibbons-Hawking boundary term is added into the action. After Euclideanization, the action and the metric can be explicitly written as
\begin{align}
S_{\epsilon} &= \int d^{n+1} x \sqrt{g} \bigg( \frac{1}{2 {\kappa_{n+1}}^2 }[R + 2 \tilde{\Lambda}] -
\frac{1}{g_v^2}[\frac{1}{4}{H}^2 + \frac{\gamma}{2}{B}^2] \bigg) + \frac{1}{{\kappa_{n+1}}^2} \int d^{n} x
\sqrt{\gamma} K,\\
{ds_{\epsilon}}^2 &= l^2 \bigg({f}(r)d \tau^2 + \frac{dr^2}{r^2} +p(r) (dx^2 + dy^2 + \cdots) \bigg),
\end{align}
where $\epsilon$ indicates the Euclidean version of the quantities.

Calculating the free energy density  (the free energy per unit $(n-1)$-dimensional spatial volume), the Einstein-Hilbert action and Gibbon-Hawking term yield
\begin{align}
&{\mathcal{F}}_{EH} = -\frac{l^{n-1}}{2 {\kappa_{n+1}}^2} \lim_{r \rightarrow 0} r \sqrt{f(r)} p'(r) p(r)^{\frac{n-3}{2}}, \label{FEDeh}\\
&{\mathcal{F}}_{GH} = \frac{1}{{\kappa_{n+1}}^2} \lim_{r \rightarrow 0} \sqrt{\gamma} K = \frac{l^{n-1}}{{\kappa_{n+1}}^2} \lim_{r \rightarrow 0} r \bigg( \sqrt{f(r)} p(r)^{\frac{n-1}{2}}
\bigg)' \label{FEDgh}
\end{align}
where $\gamma_{ab}$ is the induced metric on the boundary and $K_{\mu \nu}$ is the extrinsic curvature defined as $K_{\mu \nu} = \nabla_{\mu} n_{\nu}$ in which $n_{\nu}$ is the normal vector on the boundary surface. The free energy is   $F = \int d^{n-1} x {\mathcal{F}}$.   However for the marginally relevant modes both (\ref{FEDeh}) and (\ref{FEDgh}) are divergent  as the boundary ($r \rightarrow 0$) is approached. We incorporate boundary counterterms \cite{Ross:2009ar,Balasubramanian:2009,Mann:2011hg,Ross:2011gu} into the action as a remedy for this problem.    We construct these counterterms as a power series in
$B^2 = B^\mu B_\mu$ \cite{Cheng:2010}, so as to satisfy  covariance at the boundary, obtaining
\begin{eqnarray}
{\mathcal{F}}_{C.T.} &=& \frac{1}{2 l{\kappa_{n+1}}^2} \lim_{r \rightarrow 0} \sqrt{\gamma}
\sum^{2}_{j=0} C_{j} \bigg(-\frac{{\kappa_{n+1}}^2}{g_v^2} {B}^2 - \frac{(z-1)}{z} \bigg)^j \\
&=& \frac{l^{n-1}}{2 {\kappa_{n+1}}^2} \lim_{r \rightarrow 0} \sqrt{f(r)} p(r)^{\frac{n-1}{2}} \sum^{2}_{j=0} C_{j}
\bigg(k(r)^2 - \frac{(z-1)}{z} \bigg)^j
\end{eqnarray}
where we have used $({B}^2 - (z-1)/z)$ instead of $B^2$, since these must vanish for the pure Lifshitz case.  The coefficients
$C_{j}$  are not constants but rather a series of the logarithmic functions, with at least three needed to eliminate divergences.

The final expression for the free energy  density is
\begin{eqnarray} \label{defFED}
{\mathcal{F}} &=& {\mathcal{F}}_{EH} + {\mathcal{F}}_{GH} + {\mathcal{F}}_{C.T.} \nonumber\\
&=& \frac{l^{n-1}}{2 {\kappa_{n+1}}^2} \lim_{r \rightarrow 0} \sqrt{f(r)} p(r)^{\frac{n-1}{2}} \bigg( \frac{(n-2)r p'(r)}{p(r)}+ \frac{r f'(r)}{f(r)} + \sum^{2}_{j=0} C_{j} \bigg( k(r)^2 - \frac{(z-1)}{z} \bigg)^j  \bigg).
\end{eqnarray}

To obtain the energy density, we use the definition of the boundary stress tensor to the case in which additional non-vanishing boundary fields are present  \cite{Hollands:2005}. From the boundary stress tensor, we obtain the charge via variation of the on-shell action with respect to the boundary fields;   this process in our case produces
\begin{equation}
\delta S = \frac{\sqrt{\gamma}}{2} \tau^{ab} \delta \gamma_{ab} + {\mathcal{J}}^{a} \delta B_{a}.
\end{equation}
Here, however we are dealing not with scalar matter fields but with massive vector fields, and so the  usual charge defined by
\begin{equation}
Q = - \int d^{n-2} x \sqrt{\sigma} \xi_{a} k_{b} \tau^{ab}
\end{equation}
where $\sqrt{\sigma} = \sqrt{\gamma_{xx} \cdots \gamma_{zz}}$ is the spatial volume element, $\xi_{a}$ is a boundary Killing fields, and $k_{b}$ is the unit normal vector to the boundary Cauchy surface, is not conserved. The boundary stress tensor  $\tau^{ab}$ must therefore be redefined so as to fix the matter fields in the boundary. Employing the vielbein frame defined by
\begin{equation}
\gamma_{ab} = \eta_{\hat{a} \hat{b}} e^{\hat{a}}_{a} e^{\hat{b}}_{b}, \; \; \; \eta= \textrm{diag}(\pm1, 1, 1, \cdots)
\end{equation}
where
\begin{equation}
e^{\hat{t}} = e^{\hat{t}}_{a} dx^{a} = \sqrt{f} d \tau, \; \; e^{\hat{x_{i}}} = \sqrt{p} \; dx_{i}.
\end{equation}
We find that the variation of the free energy density retains its original form, but that $\tau^{ab}$ is replaced with
${\mathcal{T}}^{ab}$, where
\begin{equation}
\delta S = \sqrt{\gamma} {\mathcal{T}}^a_{\; \; \hat{a}} \delta e^{\hat{a}}_{a} + {\mathcal{J}}^{\hat{a}} \delta
B_{\hat{a}},
\end{equation}
with
\begin{equation}
{\mathcal{T}}^{ab} = {\mathcal{T}}^{a}_{\; \; \hat{a}} e^{b \hat{a}}, \; \; \; {\mathcal{T}}^{ab} = \tau^{ab} + \frac{1}{\sqrt{\gamma}}{\mathcal{J}}^{(a} B^{b)} .
\end{equation}
The energy density is then given by
\begin{equation}\label{endens}
{\mathcal{E}} = \sqrt{\gamma} \tau^{t}_{\; \; t} + {\mathcal{J}}^{t} B_{t}
\end{equation}
and the pressure is
\begin{equation}
{\mathcal{P}} =  - \sqrt{\gamma} \tau^{x}_{\; \;x}.
\end{equation}
Computing the distinct components of ${\mathcal{E}}$, we find
\begin{align}
\tau^{ab} &=\frac{2}{\sqrt{\gamma}} \frac{\delta {S}}{\delta \gamma_{ab}} =\frac{1}{{\kappa_{n+1}}^2} (K \gamma^{ab} - {K}^{ab}) \nonumber\\
& + \frac{1}{2 l {\kappa_{n+1}}^2} \sum^{2}_{j=0} C_{j} \bigg( \gamma^{ab} \bigg(-\frac{{\kappa_{n+1}}^2}{g_v^2} {B}^2 -\frac{(z-1)}{z} \bigg)^j  + \frac{ 2 j {\kappa_{n+1}}^2}{g_v^2} {B}^{a} {B}^{b} \bigg( -\frac{{\kappa_{n+1}}^2}{g_v^2} {B}^2 - \frac{(z-1)}{z} \bigg)^{j-1} \bigg),
\end{align}
and
\begin{eqnarray}
{\mathcal{J}}^{\hat{t}} &=& \sqrt{f(r)} \frac{\delta S }{\delta B_{t}},\nonumber\\
&=& \frac{l^{n-2}}{g_v \kappa_{n+1}} \lim_{r \rightarrow 0} \sqrt{f(r)} p(r)^{\frac{n-1}{2}} \bigg(\frac{r(k(r) \sqrt{f(r)})'}{\sqrt{f(r)}} + k(r) \sum^{2}_{j=0} j C_{j} \bigg( k(r)^2 - \frac{(z-1)}{z} \bigg)^{j-1} \bigg), \nonumber\\
&=& \frac{l^{n-2}}{g_v \kappa_{n+1}} \lim_{r \rightarrow 0} \sqrt{f(r)} p(r)^{\frac{n-1}{2}} \bigg(- \frac{1}{2}x(r) + k(r) \sum^{2}_{j=0} j C_{j} \bigg( k(r)^2 - \frac{(z-1)}{z} \bigg)^{j-1} \bigg), \label{defFlow}
\end{eqnarray}
and other component of ${\mathcal{J}}^a$ become zero. Putting these together into (\ref{endens}) yields
\begin{equation}
{\mathcal{E}} = \frac{l^{n-1}}{2 {\kappa_{n+1}}^2} \lim_{r \rightarrow 0} \sqrt{f(r)} p(r)^{\frac{n-1}{2}} \bigg( \frac{(n-1)r p'(r)}{p(r)} -  x(r) k(r) + \sum^{2}_{j=0} C_{j} \bigg( k(r)^2 -\frac{(z-1)}{z} \bigg)^j  \bigg) \label{defED}
\end{equation}
 and
\begin{equation}{\label{pressure}}
{\mathcal{P}} = - {\mathcal{F}}.
\end{equation}
Imposing finiteness of physical quantities of (\ref{defFED}), (\ref{defFlow}), and (\ref{defED}), the coefficients of the counter terms are found to be
\begin{align}
C_{0} =& 2(2z-1) - \frac{2z^2}{(2z-1)(z-1)^3 \log^2(r \Lambda)} + \frac{(4z^4 + 2z^3 -3z^2-2z+1)}{(z-1)^5(2z-1)^2 \log^3(r \Lambda)} \label{C0} \nonumber\\
& +\frac{4z(3z-1) \log(-\log(r \Lambda))}{(z-1)^5(2z-1) \log^3(r \Lambda)}+ \cdots, \\
C_{1} =& z + \frac{2z^3}{(2z-1)(z-1)^2 \log(r \Lambda)} - \frac{z(14z^3 - z^2 - 10z + 3) + 4z^2(2z-1)(3z-1)\log(-\log(r \Lambda))}{2(2z-1)^2(z-1)^4 \log^2 (r \Lambda)} \nonumber\\
& - \frac{1}{\log^3(r \Lambda)} \bigg( \frac{z^2 (34z^4 -54z^3 +72z^2 -46z+9) + 8 a (2z-1)^2(z-1)^5}{2 z (2z-1)^2(z-1)^6} \nonumber\\
& - \frac{(6z^4 +25z^3 -45z^2+21z-3) \log(-\log(r \Lambda))}{(2z-1)^2(z-1)^6} - \frac{2z(3z-1)^2 \log^2 (-\log(r \Lambda))}{(z-1)^6(2z-1)} \bigg) + \cdots, \label{C1} \\
C_{2} =& \frac{z^2(1-3z)}{4(2z-1)(z-1)} + \frac{z^2(15z^2-14z+3)}{4(2z-1)^2(z-1)^3 \log(r \Lambda)} + \frac{a}{\log^2 (r \Lambda)} - \frac{z(3z-1)^2(5z-3)\log(-\log(r \Lambda))}{4(2z-1)^2 (z-1)^5 \log^2 (r \Lambda)} \label{C2}
\end{align}
where the first two are  infinite series in $1/\log(r \Lambda)$ that include powers of $\log(-\log(r \Lambda))$ such that the order of the $\log(-\log(r \Lambda))$ terms do not exceed the order of the $1/\log(r \Lambda)$ terms. It is sufficient for $C_2$ to retain
terms up to   second order in  $1/\log(r \Lambda)$.  Note that there exists an ambiguity $a$ in these expressions. This ambiguity does not affect   numerical evaluation of the free energy density and the energy density that we shall later compute, though it does
affect ${\mathcal J}^{\hat{t}}$, reflecting the reaction of the system to changes in the boundary Proca field.
Our counter term  construction  (\ref{C0}) -- (\ref{C2}) is minimal; additional terms such as
$C_3 (B^2 - (z-1)/z)^3 $ or $C_4 (B^2 - (z-1)/z)^4$ would also yield solutions.

Applying (\ref{C0}) -- (\ref{C2}) into (\ref{defFED}), (\ref{defFlow}) and (\ref{defED}), the physical quantities become
\begin{align}
&{\mathcal{F}} = \frac{l^{n-1}}{{\kappa_{n+1}}^2} \frac{\sqrt{z}}{\sqrt{z-1}(2z-1)} \bigg(z \alpha - \frac{(2z^3-2z^2-2z+1)}{(2z-1)(z-1)^4} \beta \bigg), \label{asympFED}\\
&{\mathcal{E}} = - \frac{l^{n-1}}{{\kappa_{n+1}}^2} \frac{\sqrt{z}}{\sqrt{z-1}(2z-1)} \bigg( z \alpha + \frac{(2z^3-4z^2+4z-1)}{(2z-1)(z-1)^4}\beta \bigg), \label{AsympED}\\
&{\mathcal J}^{\hat{t}} = \frac{1}{g_v} \frac{l^{n-2}}{\kappa_{n+1}} \bigg(\frac{z(20z^5 + 18z^4-22z^3 -23z^2+24z-5)}{2(z-1)^4(2z-1)^3} + \frac{4(z-1)a}{z} \bigg)\beta.
\end{align}
Since the pure Lifshitz solution does not depend on $\alpha$ and $\beta$,  we expect in this case that
\begin{equation}
{\mathcal{F}} = {\mathcal{E}} = {\mathcal J}^{\hat{t}} = 0
\end{equation}
regardless of the dimension of  spacetime.

\section{Finite Temperature}
\label{HD expn}

 In this section, we consider the finite temperature theory by expanding the black hole solution near horizon. Our goal is  to describe the renormalization group (RG) flow under the marginally relevant modes, and to predict behaviour  of  physical quantities such as the free energy density $\mathcal{F}$ and the energy density $\mathcal{E}$   near  criticality in order to provide a way of understanding the quantum phase transition from one phase to the critical point via the holographic dictionary. For these purposes, the RG flow corresponding to the $h_0$-dependent horizon flux    is explained by ensuring the RG flow in the  zero temperature limit {\cite{Kachru:2008}} $\Lambda^z/T \rightarrow \infty$ with $\Lambda \sim 0$ fixed. Furthermore ${\mathcal{F}}/Ts$ and ${\mathcal{E}}/Ts$ are found as functions of $\log(\Lambda^z/T)$ in the near-Lifshitz spacetime containing the large flux.

\subsection{Expansion and Physical quantities near horizon}
\label{sec:NearHrz}

We assume a black hole solution having the form of (\ref{metanz}) and defined by $f(r_+)=0$, and expand the solution near horizon $r=r_+$. In this expansion the unknown coefficients arising in every order in the $f(r),g(r)$ and $h(r)$ are calculated by the equations of the motion (\ref{newvars}). Regularity requires   $g_{tt}$  to have a double zero at the horizon, with $g_{xx}$ remaining nonzero. Under these considerations the functions are expressed by
\begin{align}
&f(r)=f_{0} \bigg( \bigg(1-\frac{r}{r_{+}} \bigg)^2 + \bigg(1-\frac{r}{r_{+}} \bigg)^3 + \frac{(-6z^2+14z+7)z+8(3z-2)h_{0}^2 }{12z}\bigg(1-\frac{r}{r_{+}} \bigg)^4 + \cdots \bigg), \label{nhF}\\
&p(r)=p_{0} \bigg(1 + \frac{(3 z-1)z- 4 h_{0}^2}{2z} \bigg(1-\frac{r}{r_{+}} \bigg)^2 + \frac{(3 z-1)z- 4 h_{0}^2}{2z}
\bigg(1-\frac{r}{r_{+}} \bigg)^3 + \cdots  \bigg), \label{nhP}\\
&h(r)=\sqrt{f_{0}} \bigg( h_{0} \bigg(1-\frac{r}{r_{+}} \bigg)^2 + h_{0} \bigg(1-\frac{r}{r_{+}} \bigg)^3 + h_{0}
\bigg(\frac{z(-9z^2+10z+20)+ 8 h_{0}^2 (3z-1)}{24z}\bigg) \bigg(1-\frac{r}{r_{+}} \bigg)^4  + \cdots\bigg) \label{nhH}
\end{align}
where the constants $f_{0}$ and $p_{0}$ are associated with scaling ambiguities of the coordinates $\{t, x, y, \cdots \}$, which function
as the clock and rulers of the system. Upon fixing these the only variable left in the metric is $h_0$.  Different values of  $h_0$ correspond to different black holes and so we have a 1-parameter  family of black hole  solutions.

Some physical information, for example thermodynamic  quantities, can be obtained near the horizon of the black hole. At $r=r_+$, the temperature  $T$, can be computed by identifying the imaginary time coordinate $\tau$ with period $\beta$ so as to ensure regularity of the metric, and the entropy density, $s$, obtained from the definition of the entropy, $S = \frac{A}{4 G_{n+1}}$. We find
\begin{align}
&T = \frac{r_{+}}{2 \pi} \sqrt{\frac{1}{2} \frac{d^2 f(r)}{dr^2}} \bigg|_{r=r_{+}} ,\\
&s = 2 \pi \frac{l^{n-1}}{{\kappa_{n+1}}^2} p(r_{+})^{\frac{n-1}{2}},
\end{align}
where $S=\int s \; d^{n-1} x $.

The horizon flux, $\Phi$, of the massive vector field is
\begin{equation}
\Phi = \oint \sqrt{h} \vec{E} \cdot d \vec{A} = \oint \phi \; d^{n-1}x
\end{equation}
where
\begin{equation}
\phi = \frac{l^{n-2} g_v r_{+}}{\kappa_{n+1}} \bigg( \frac{p(r)^{\frac{n-1}{2}}}{\sqrt{f(r)}} \frac{dh(r)}{dr} \bigg) \bigg|_{r=r_{+}}
\end{equation}
is the horizon flux density.
Using (\ref{nhF}) --(\ref{nhH}) we obtain
\begin{equation}
T = \frac{\sqrt{f_{0}}}{2 \pi}, \; \; \; \; s = 2 \pi p_{0}^{\frac{n-1}{2}} \frac{l^{n-1}}{{\kappa_{n+1}}^2}, \; \; \; \; \phi = 2 h_{0}p_{0}^{\frac{n-1}{2}} \bigg( \frac{l^{n-2} g_v}{\kappa_{n+1}}\bigg). \label{thermoVar}
\end{equation}
for the temperature,   entropy density, and horizon flux density.

\subsection{Integrated First law of thermodynamics}
\label{sec:NearHrz}

Before embarking on our numerical calculations, in this section we obtain relationships between the  free energy density and the energy density  derived in section 4 and the above thermodynamic quantities. We then use this to obtain
 analytic predictions when the marginally relevant mode vanishes, i.e. $\Lambda \rightarrow 0$.

First, by using the asymptotic solutions (\ref{AsympSolk}) -- (\ref{AsympSolx}) we construct an $r$-independent
RG-invariant quantity
\begin{align}
\bar{K} &= -\frac{1}{2} \sqrt{f(r)} p(r)^{\frac{n-1}{2}} \bigg( -q(r) + m(r) + k(r) x(r) \bigg), \nonumber\\
&= - \frac{\sqrt{f(r)} p(r)^{\frac{n-1}{2}}}{4 (n-1) q(r)} \bigg( 4 \chi + 4(n-1)z k(r)^2 - n(n-1)q(r)^2 - x(r)^2 + 2(n-1)q(r)k(r)x(r) \bigg),
\end{align}
where plugging (\ref{HgExpk})--(\ref{HgExpx})
into the above gives
\begin{equation}
\bar{K}= \frac{2 \sqrt{z}}{\sqrt{z-1}(1-2z)} \bigg( z \alpha - \frac{(z^2 - 3z+1)}{(1-2z)(z-1)^2} \beta \bigg).
\end{equation}
Near the horizon, this RG-invariant quantity can be calculated by using (\ref{nhF}) --(\ref{nhH}) and expressed in terms of $T$ and $s$ by applying (\ref{thermoVar}). We find
\begin{equation} \label{KintermsofTs}
\bar{K} = \sqrt{f_{0}} p_{0}^{\frac{n-1}{2}} = Ts \frac{{\kappa_{n+1}}^2}{l^{n-1}}.
\end{equation}
Next, from the free energy density (\ref{defFED}) and the energy density (\ref{defED}) we obtain the following relation
\begin{align}
&\frac{1}{2} \sqrt{f(r)} p(r)^{\frac{n-1}{2}} \bigg( \frac{(n-1)r p'(r)}{ p(r)} -  x(r) k(r) + \sum^{2}_{j=0} C_{j} \bigg( k(r)^2
-\frac{(z-1)}{z} \bigg)^j  \bigg) \nonumber\\
&= \frac{1}{2}  \sqrt{f(r)} p(r)^{\frac{n-1}{2}} \bigg( \frac{(n-2)r p'(r)}{p(r)} + \frac{r
f'(r)}{f(r)} + \sum^{2}_{j=0} C_{j} \bigg( k(r)^2 - \frac{(z-1)}{z} \bigg)^j  \bigg) + \bar{K},
\end{align}
which is more simply expressed as
\begin{equation} \label{PreFrstThlaw}
{\mathcal{E}} = {\mathcal{F}} + \frac{l^{n-1}}{{\kappa_{n+1}}^2} \bar{K}.
\end{equation}
This relation is easily checked using (\ref{asympFED}), (\ref{AsympED}) and (\ref{KintermsofTs}).

Finally, combining (\ref{KintermsofTs}) with (\ref{PreFrstThlaw}) gives
\begin{equation}
{\mathcal{F}} = {\mathcal{E}} - T s. \label{FirThLaw}
\end{equation}
which is the integrated form of the first law for these black holes. We will use this to check the accuracy of our numerical results in section 5.4.

Considering the limit $\Lambda = 0$, since anisotropic scale invariance  still holds, from the Ward identity we expect  that the pressure is equal to the energy \cite{Taylor:2008}. From (\ref{pressure}), we have
\begin{equation}
{\mathcal{F}}_{0} = - {\mathcal{E}}_{0} \; \; \; \; \; \; \; (\Lambda = 0),
\end{equation}
and in conjunction with (\ref{FirThLaw}), we obtain an analytic prediction for when the marginally relevant modes are not excited
\begin{equation}
{\mathcal{F}}_{0} = - {\mathcal{E}}_{0} = -\frac{1}{2} T s_{0} \; \; \; \; (\Lambda = 0),
\end{equation}
which will be used for a consistency check on our numerical results in section 5.4. Note that the relation ${\mathcal{F}} = - {\mathcal{E}}$ also holds for $\Lambda \sim 0$ case when $\beta = 0$, as is easily seen from equations (\ref{asympFED}) and (\ref{AsympED}).

\subsection{Integrating towards the Lifshitz Boundary}
\label{sec:NearHrz}

To investigate Lifshitz spacetime in the UV-region, $T \gg \Lambda^z$,   we re-expand the asymptotic solutions into the high energy regime by using (\ref{Expsn}), and set the arbitrary scale $\mu \sim r_{+}^{-1}$, thereby making the marginally relevant modes  near-Lifshitz. Upon carrying this out, the deformed Lifshitz spacetime is described in the high temperature regime with fixed $\Lambda \sim 0$. This spacetime approaches a  pure-Lifshitz by supplying heat  at much greater temperatures or higher energy scales.
This process of expansion yields a constant $\log(\Lambda r_+)$ which must be fixed in the asymptotic regions. Near the horizon  the expanded black hole solution has three parameters $f_0$, $p_0$, and $h_0$ where $f_0$ and $p_0$ will be determined, depending on which asymptotic spacetime is imposed.

In this section, we numerically connect the expanded near-horizon black hole solution with the near-Lifshitz asymptotic solution. By matching two solutions, we determine the constant $\log(\Lambda r_+)$ which corresponds to $h_0$ in section 5.3.1, and $f_0$ and $p_0$ in section 5.3.2. Then with these values we compute the thermodynamic quantities, varying the value of $h_0$. We compute the entropy density as a function of $\log(\Lambda^z/T)$   in section 5.3.2, and the free energy density and energy density versus $\log(r/r_+)$  in section 5.3.3. In section 5.4, we compute the free energy density and energy density as functions of $\log(\Lambda^z/T)$ and find a suitable fitting curve, obtaining a prediction on the sub-leading order of the free energy and energy density as a function of $\log(\Lambda^z/T)$.  We also discuss how   renormalization group flow is described in our context in section 5.3.1.  In our numerical work, we use $r/r_+$ as our radial variable, and unitless quantities such as ${\mathcal{F}}/Ts$, ${\mathcal{E}}/Ts$, and $\Lambda^z/T$ are considered.


\subsubsection{Matching $\Lambda$}
\label{sec:NearHrz}
\begin{figure}[p]
        \subfloat[ $z=2$ and $h_0 = 0.97128$, $\log(\Lambda r_{+})=-11141.7$]{\includegraphics[scale=0.8]{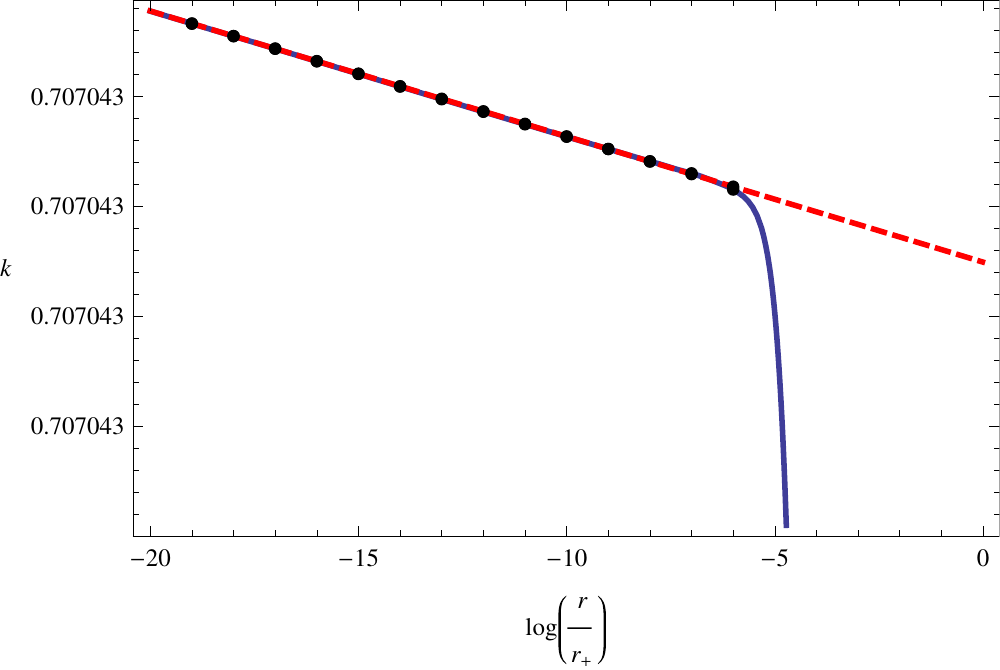}} \; \; \;
        \; \;
        \subfloat[ $z=3$ and $h_0 = 1.63428$, $\log(\Lambda r_{+})=-4188.2$]{\includegraphics[scale=0.8]{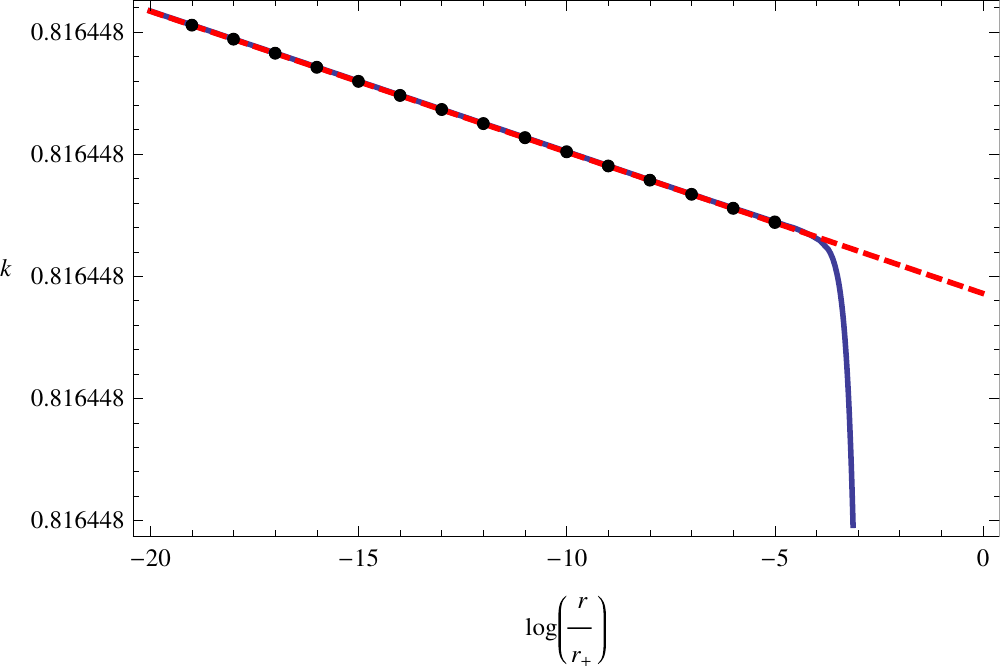}}\\
        \vspace{15pt}\\
        \subfloat[ $z=4$ and $h_0 = 2.28218$, $\log(\Lambda r_{+})=-2709.4$]{\includegraphics[scale=0.8]{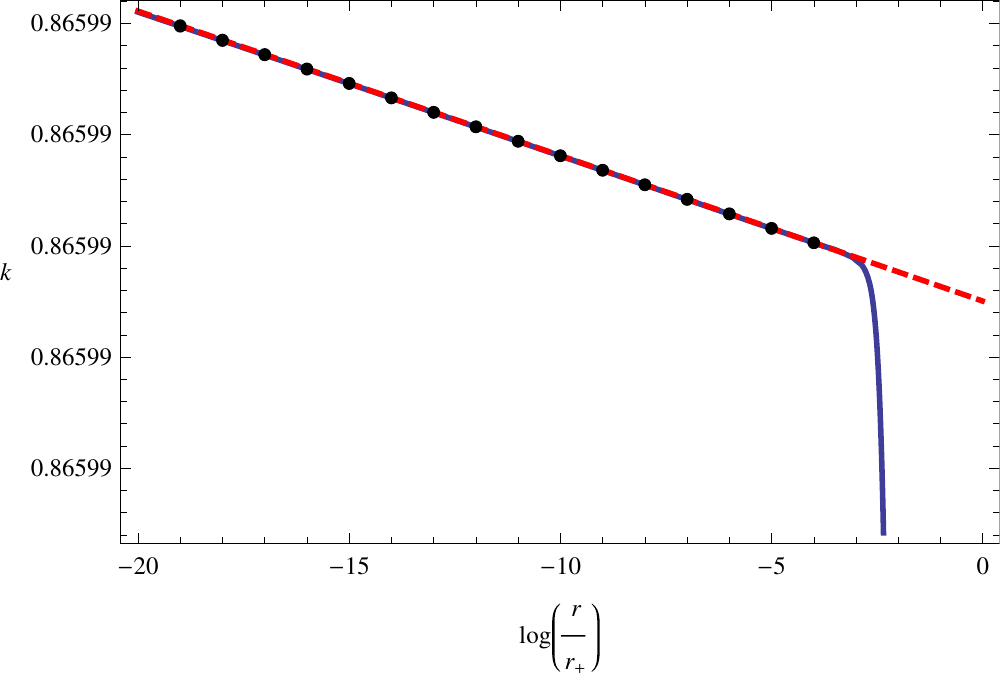}} \; \; \;
        \; \;
        \subfloat[ $z=5$ and $h_0 = 2.92548$, $\log(\Lambda r_{+})=-2270.6$]{\includegraphics[scale=0.8]{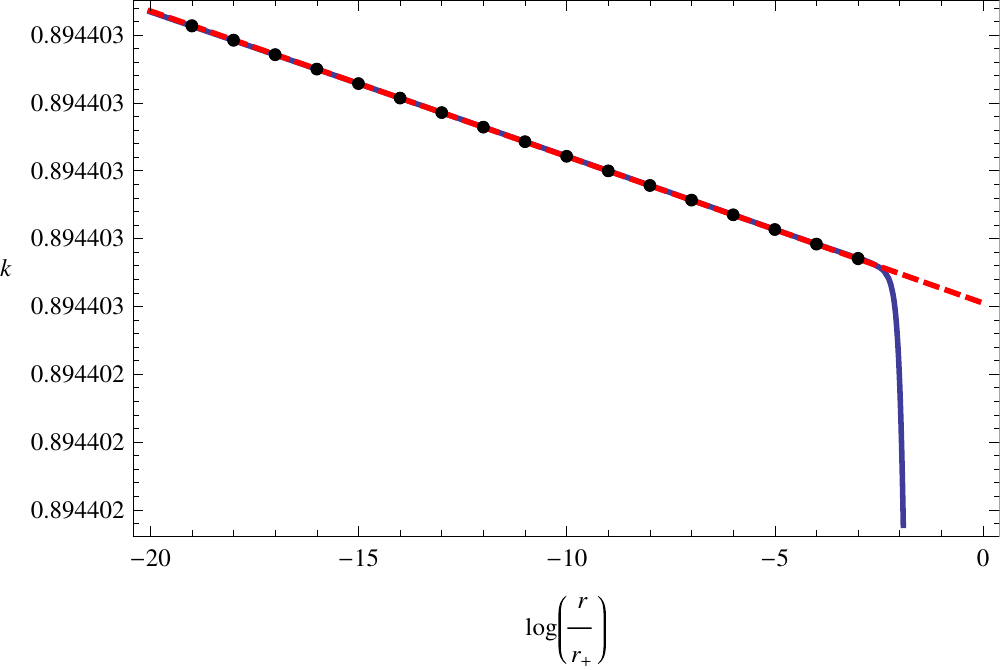}}\\            \vspace{15pt}\\
        \subfloat[ $z=6$ and $h_0 = 3.56678$, $\log(\Lambda r_{+})=-2726$]{\includegraphics[scale=0.8]{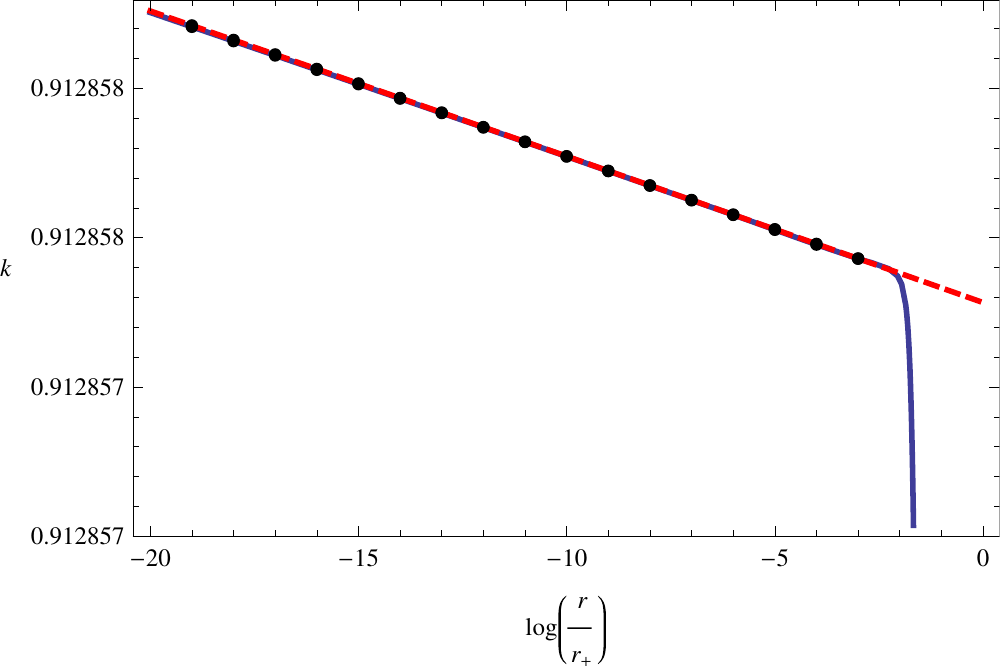}} \; \; \; \;
        \;
        \subfloat[ $z=7$ and $h_0 = 4.20680$, $\log(\Lambda r_{+})=-908.7$]{\includegraphics[scale=0.8]{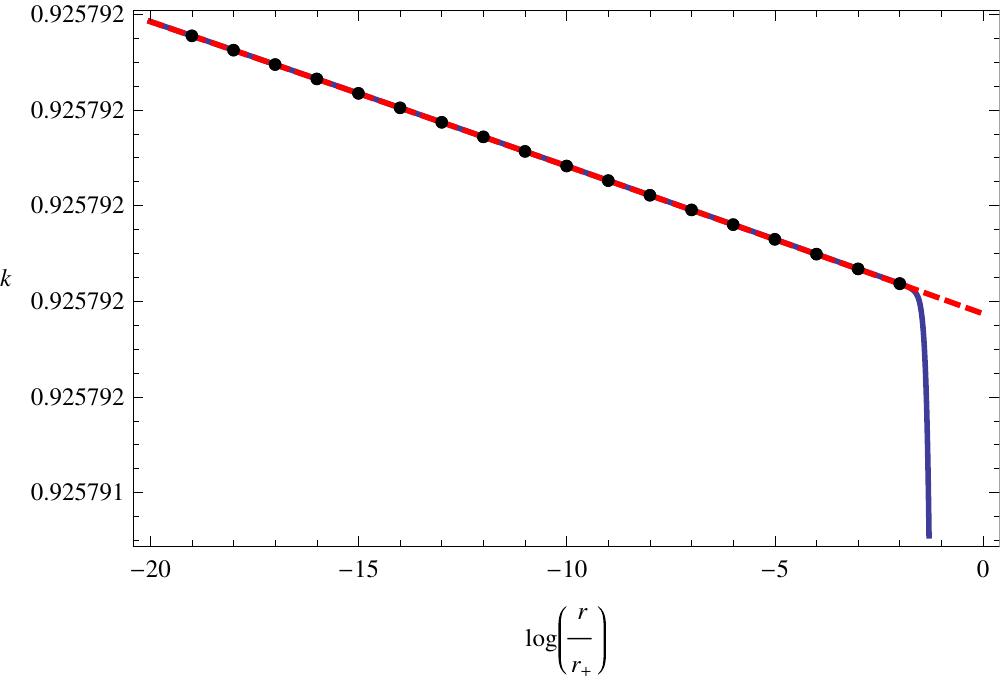}}               \caption{Extracting $\log( \Lambda r_{+})$}
        \label{fig:extLam}
\end{figure}

For extracting $\Lambda$, we start with the asymptotic solutions $k(r)$, $q(r)$, and $x(r)$ in (\ref{AsympSolk}) -- (\ref{AsympSolx}). As the UV-deformation is applicable to  Lifshitz spacetime in the high energy limit $\Lambda^z/T \rightarrow 0$, we expand the solutions as in (\ref{HgExpk}) -- (\ref{HgExpx}),  bringing in the arbitrary scale $\mu >> \Lambda$ eventually setting $\mu \to r_{+}^{-1}$.
Then, near the asymptotic boundary, we have  only to fix the variable $\log(\Lambda r_+)$.  Near the horizon, the expanded black hole solutions (\ref{nhF})--(\ref{nhH}) transfer to functions of $k(r)$, $q(r)$, and $x(r)$ by (\ref{newvars}) and (\ref{newx}), and in this process $f_0$ and $p_0$ drop out and $h_0$ only remains. Now, having a proper value of $h_0$, we numerically integrate from the near horizon to the boundary satisfying the equations of motion (\ref{eqkqx}),  matching the numerical solutions to the asymptotic expectation by adjusting the value of $\log(\Lambda r_+)$ in the middle region. In our numerical works, the integration  starts at $\log(r/r_+) \sim -0.015$ and ends at $\log(r/r_+) \sim -10^{4}$. Our numerical results for the $k$ function are shown in Figure~\ref{fig:extLam}, where the red dashed line is a fit for the asymptotic expectation, the blue solid line corresponds to the numerical results, and   the dots signify the values used for finding the matching condition.  We note that the agreement between our numerical results and the asymptotic expectations is very strong.

\begin{figure}[t!]
    \begin{center}
    \includegraphics{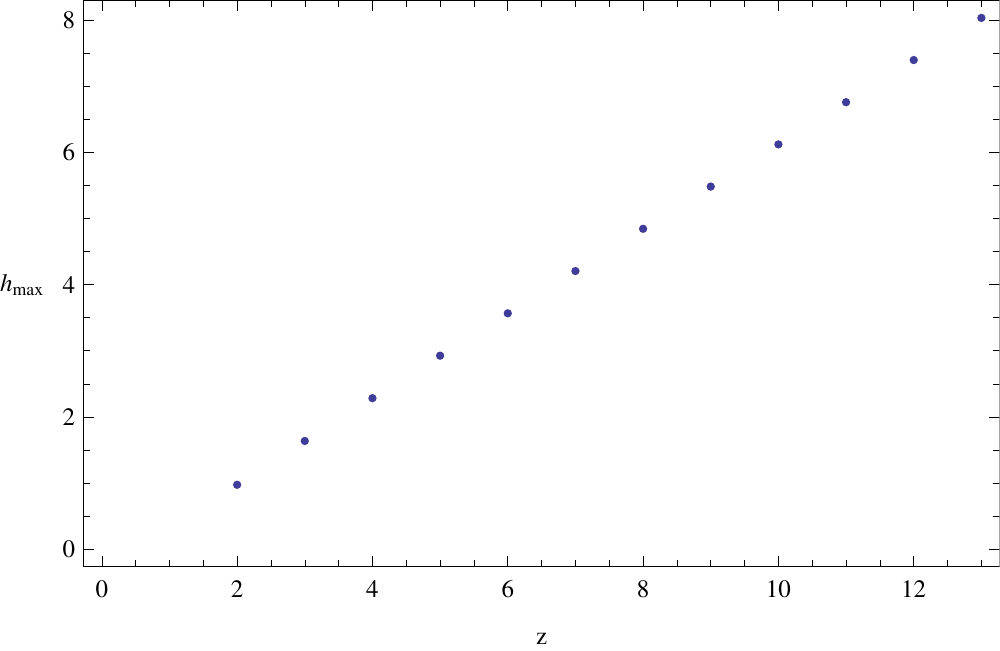}
    \caption{$h_{max}$ versus $z$}
    \label{fig:hmax}
    \end{center}
\end{figure}

\begin{table}[b!]
\begin{center}
  \begin{tabular}{| c || c |}
    \hline
      & $\; \; \; \; \; \; \; \; \; \; \; \; \; \; \; \; \;h_{max}\; \; \; \; \; \; \; \; \; \; \; \; \; \; \; \;$ \\ \hline
    $z=2$ & 0.9713 \\ \hline
    $z=3$ & 1.6343 \\ \hline
    $z=4$ & 2.2822 \\ \hline
    $z=5$ & 2.9255 \\ \hline
    $z=6$ & 3.5668 \\ \hline
    $z=7$ & 4.2070 \\ \hline
    $z=8$ & 4.8465 \\ \hline
    $z=9$ & 5.4856 \\ \hline
    $z=10$ & 6.1244 \\ \hline
    $z=11$ & 6.7629 \\ \hline
    $z=12$ & 7.4012 \\ \hline
    $z=13$ & 8.0394 \\ \hline
    $\vdots$ & $\vdots$ \\
    \hline
    \end{tabular}
    \caption{maximum value of $h_0$}
    \label{table:hmax}
  \end{center}
\end{table}

The important outcome of this procedure is that a maximum value of $h_0$ is obtained. In other words, if we keep increasing the value of $h_0$ then beyond a certain point we are not able to find a matching condition connecting our numerical result to the asymptotic expectation. Physically this means that at a large value of flux, the metric functions grow exponentially as the boundary is approached, so they do not ever reach the boundary. This means that under the condition of  large flux and  high temperature $\Lambda^z/T \rightarrow 0 $, the spacetime having the black hole (\ref{nhF})--(\ref{nhH}) is no longer deformed, but rather is asymptotic to pure Lifshitz spacetime. We also find that the maximum values of $h_0$  linearly increased according to the critical exponent, $z$ (or the spatial dimension) of the spacetime, $n$. We present  this behaviour in Figure~\ref{fig:hmax}, and explicitly denote the maximum values of $h_0$ in Table {~\ref{table:hmax}}.

On the other hand, a minimum value of the flux also exists and its value becomes zero at the horizon. In this case, the massive vector field disappears and the spacetime is described by an asymptotically AdS black hole. Examining the situation for
 $h_0$   between $h_{max}$ and $h_{min}$ is non-trivial.  However for a small amount of flux  we can verify that
in the zero temperature limit $\Lambda^z/T \rightarrow \infty$  the renomalization group flow is recovered {\cite{Kachru:2008}}.   This obviously indicates the existence of an RG flow with the marginally relevant mode, and also implies that tuning the horizon flux via $h_0$ interpolates between the zero temperature RG flow {\cite{Kachru:2008}} and   asymptotic  Lifshitz black holes
{\cite{Ross:2009ar,Taylor:2008,Bertoldi:2005,Mann:2009,Danielsson:2009,Bertoldi:2009}}.

\subsubsection{Matching $f_{0}$, and $p_{0}$}
\label{sec:NearHrz}

\begin{figure}[p]
        \subfloat[For $z=2$ and $h_0 = 0.962$, which corresponds to $\log(\Lambda r_{+})=-106.8$, the red dashed line is matched to the blue solid line at $f_{0} r_{+}^4 = 29.04$ (left) and at $p_{0} r_{+}^2 = 1.8428$ (right).  The dots correspond to the values used for finding the matching condition.]{\includegraphics[scale=0.8]{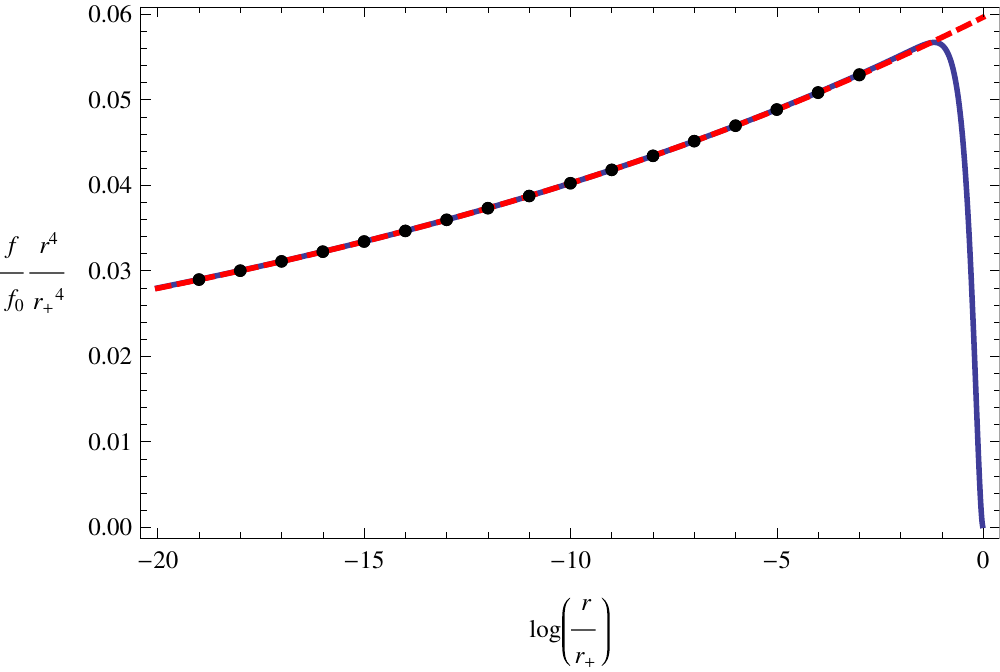} \; \; \; \; \; \includegraphics[scale=0.8]{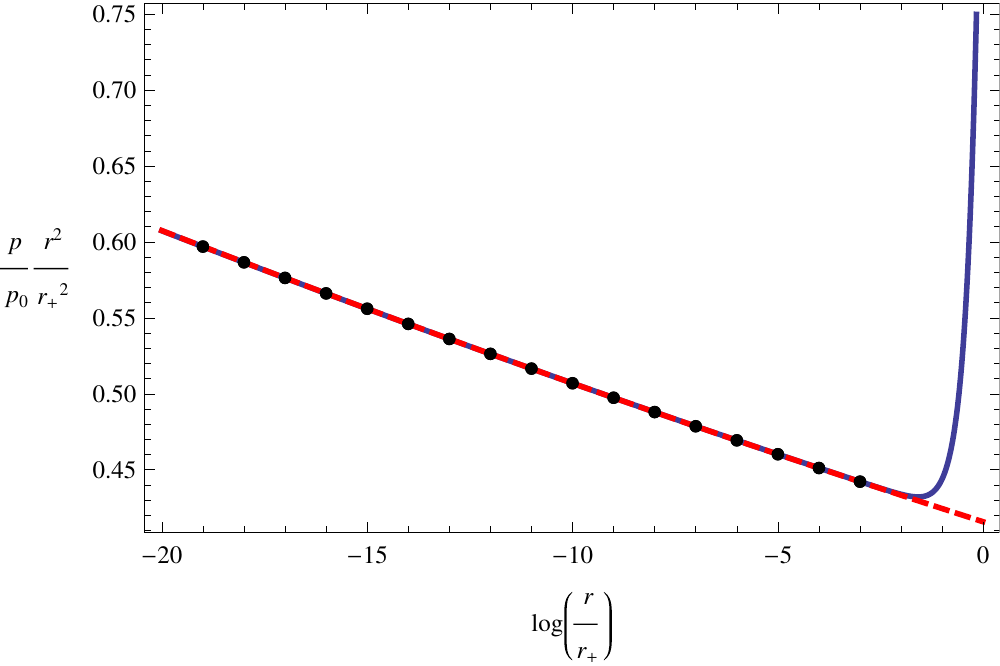}} \\
        \vspace{10pt}\\
        \subfloat[For $z=3$ and $h_0 = 1.628$, which corresponds to $\log(\Lambda r_{+})=-72.6$, the red dashed line is matched to the blue solid line at $f_{0} r_{+}^6 = 44.266$ (left) and at $p_{0} r_{+}^2 = 1.7579$ (right). The dots correspond to the values used for finding the matching condition.]{\includegraphics[scale=0.8]{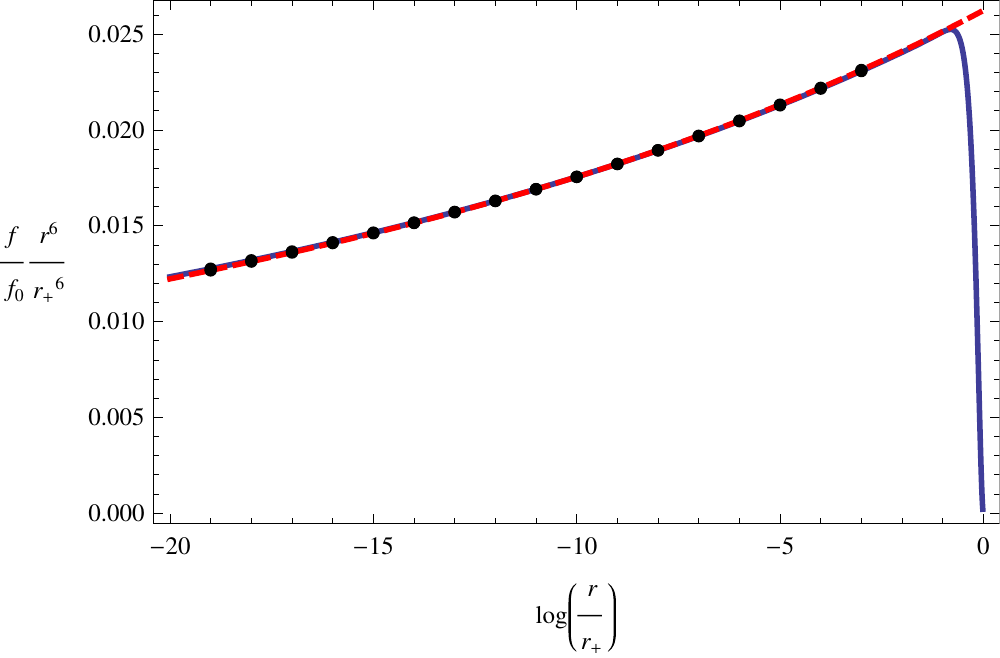} \; \; \; \; \; \includegraphics[scale=0.8]{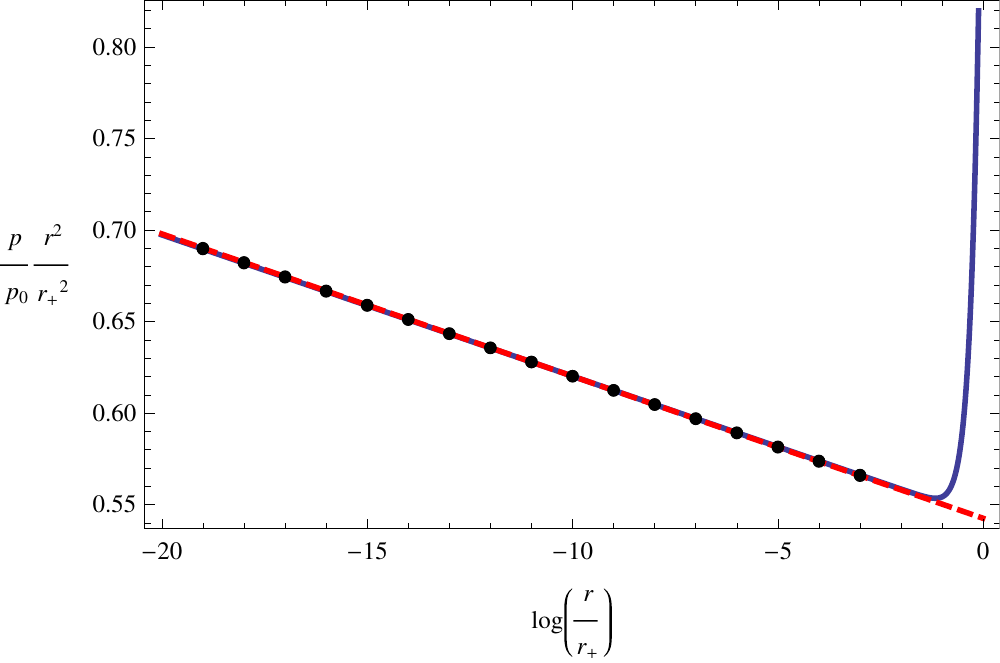}} \\
        \vspace{10pt}\\
        \subfloat[For $z=4$ and $h_0 = 2.2798$, which corresponds to $\log(\Lambda r_{+})=-133.4$, the red dashed line is matched to the blue solid line at $f_{0} r_{+}^8 = 71.40$ (left) and at $p_{0} r_{+}^2 = 1.5781$ (right).  The dots correspond to the values used for finding the matching condition.]{\includegraphics[scale=0.8]{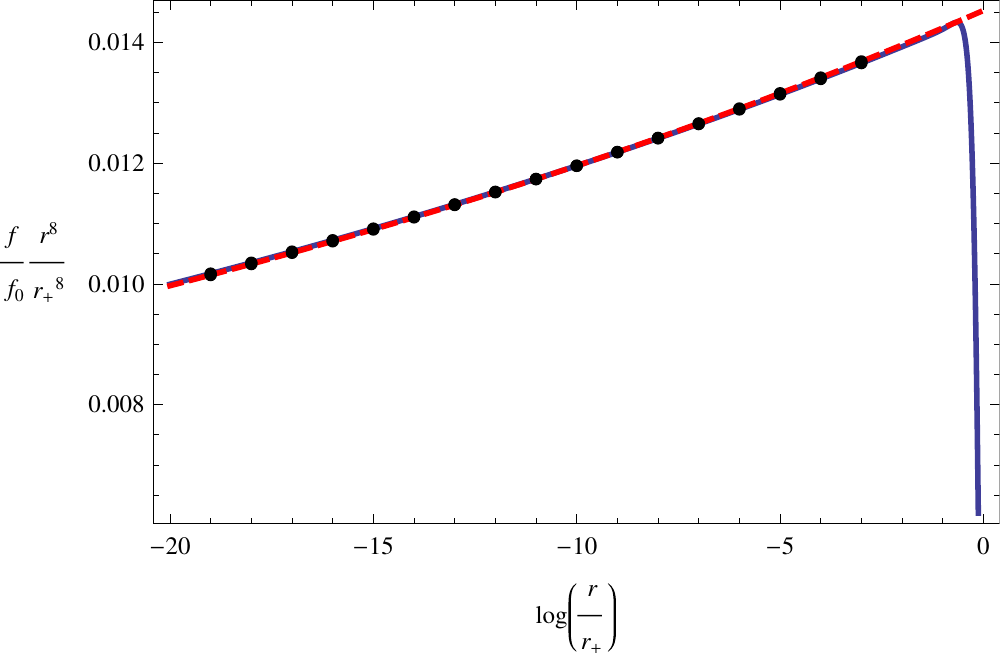} \; \; \; \; \; \includegraphics[scale=0.8]{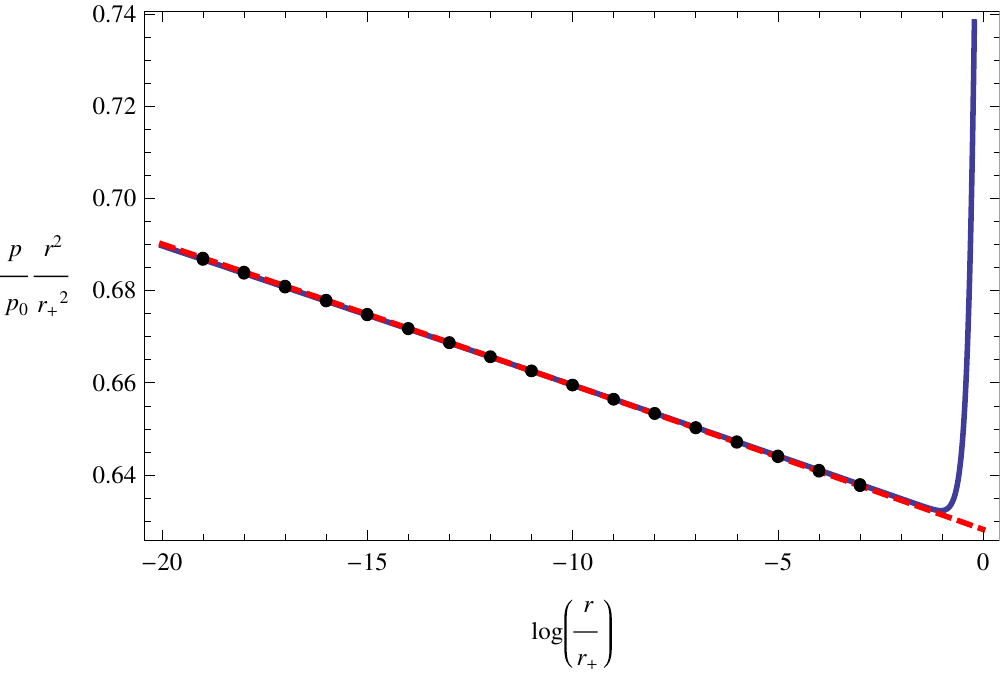}} \\
        \caption{Extracting $f_{0}$ and $p_{0}$ for $z=2,3,$ and $4$}
        \label{fig:extfandp1}
\end{figure}

\begin{figure}[p]
        \subfloat[For $z=5$ and $h_0 = 2.9238$, which corresponds to $\log(\Lambda r_{+})=-155.6$, the red dashed line is matched to the blue solid line at $f_{0} r_{+}^{10} = 109.953$ (left) and at $p_{0} r_{+}^2 = 1.45079$ (right). The dots correspond to the values used for finding the matching condition.]{\includegraphics[scale=0.8]{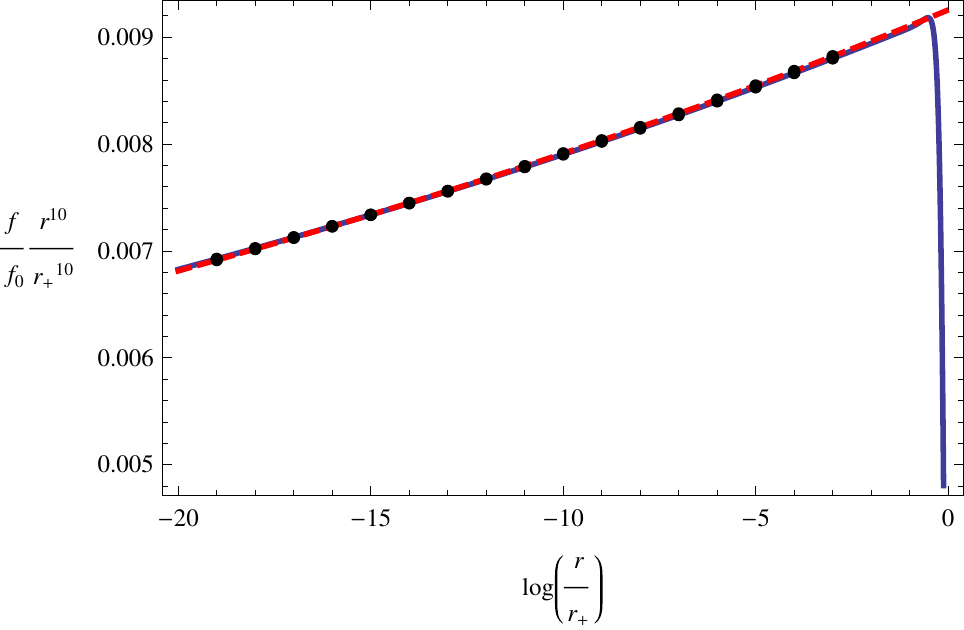} \; \; \; \; \; \includegraphics[scale=0.8]{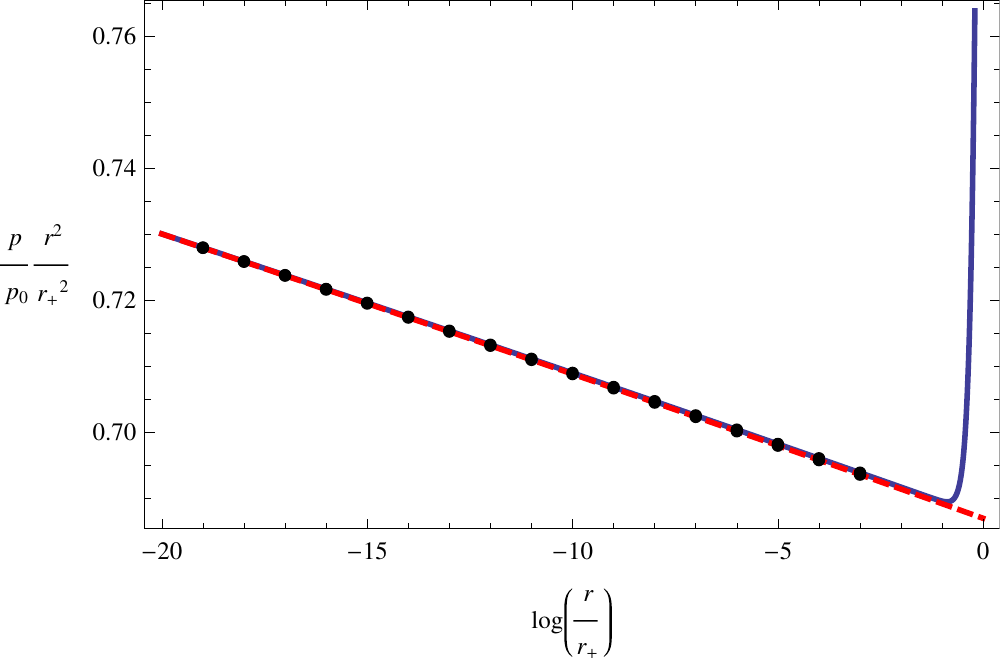}} \\
        \vspace{10pt}\\
        \subfloat[For $z=6$ and $h_0 = 3.5655$, which corresponds to $\log(\Lambda r_{+})=-183$, the red dashed line is matched to the blue solid line at $f_{0} r_{+}^{12} = 157.777$ (left) and at $p_{0} r_{+}^2 = 1.36751$ (right). The dots correspond to the values used for finding the matching condition.]{\includegraphics[scale=0.8]{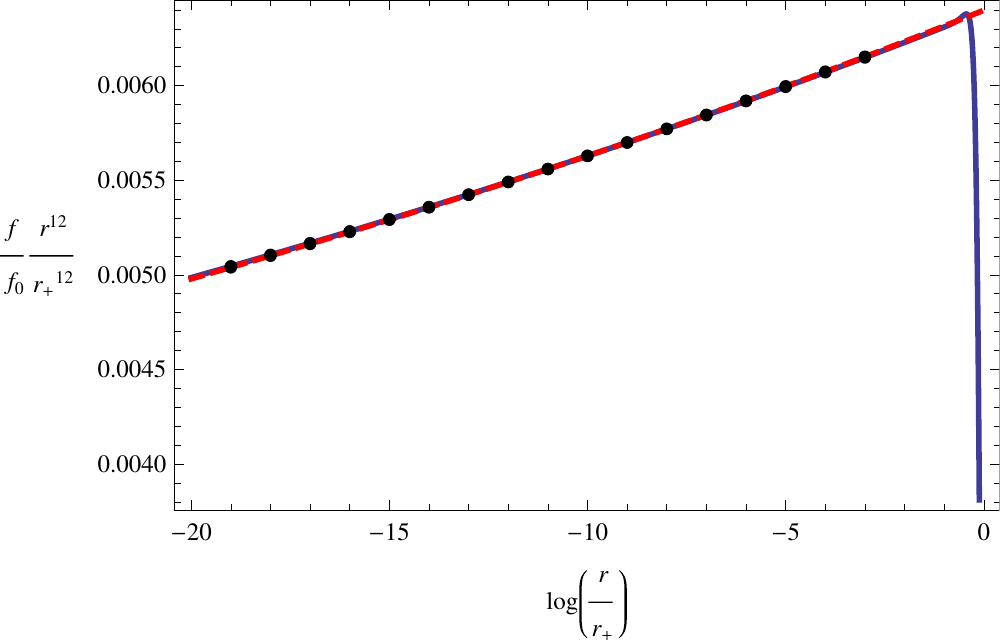} \; \; \; \; \; \includegraphics[scale=0.8]{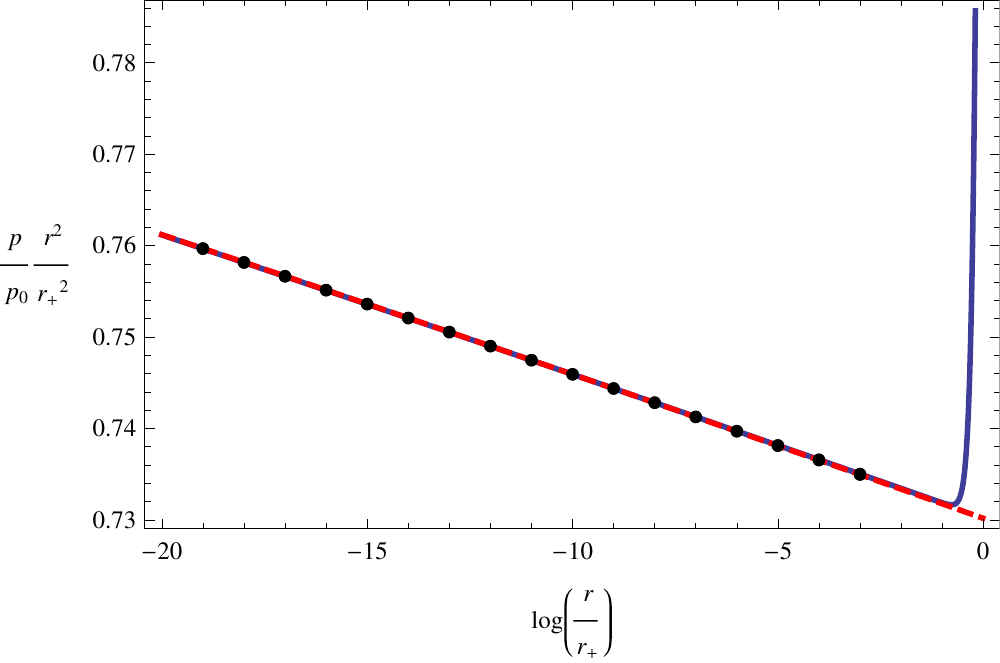}} \\
        \vspace{10pt}\\
        \subfloat[For $z=7$ and $h_0 = 4.206$, which corresponds to $\log(\Lambda r_{+})=-220.2$, the red dashed line is matched to the blue solid line at $f_{0} r_{+}^{14} = 214.442$ (left) and at $p_{0} r_{+}^2 = 1.3098$ (right). The dots correspond to the values used for finding the matching condition.]{\includegraphics[scale=0.8]{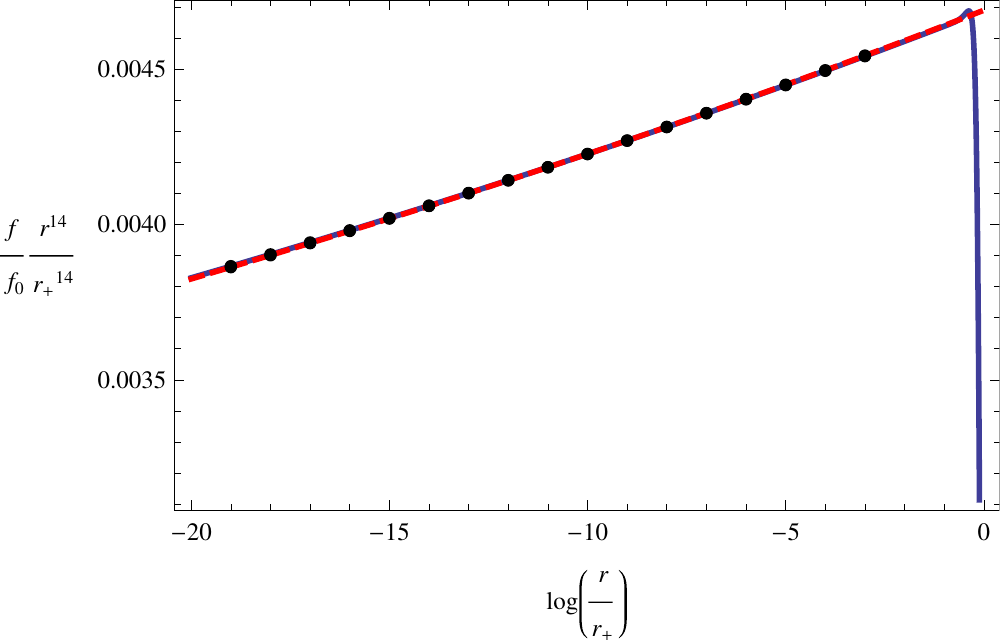} \; \; \; \; \; \includegraphics[scale=0.8]{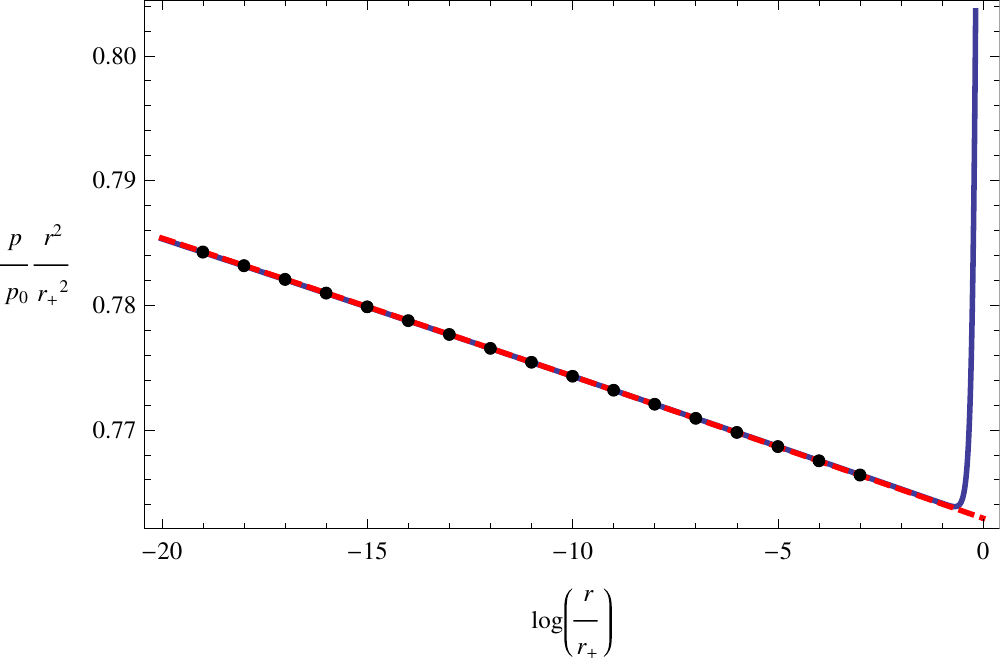}} \\
        \caption{Extracting $f_{0}$ and $p_{0}$ for $z=5,6,$ and $7$}
        \label{fig:extfandp2}
\end{figure}

Since, given $h_0$, $\log(\Lambda r_+)$   is fixed, $f_0$ and $p_0$ arising in (\ref{nhF}) -- (\ref{nhH}) can be determined. Repeating the previous process with the functions of $f$, $p$, and $h$, the asymptotic solutions (\ref{AsympSolf}) -- (\ref{AsympSolp}) are in the high energy regime obtained via the expansion of (\ref{Expsn}) and the rescaling of (\ref{scalingtx}), becoming the expressions (\ref{hgexpf})--(\ref{hgexpp}).  We can neglect the $\alpha$ and $\beta$ involved in the exponent terms because these  contribute much less than the   logarithmic terms as the boundary is approached (technically these unknowns  are hard to  fix). Applying the same numerical technique, we start to integrate from the near horizon towards the boundary governed by the equation of motion (\ref{eqfph}). Over a finite range, provided values of $h_0$ and $\log(\Lambda r_{+})$, the numerical solutions and  asymptotic expectations (\ref{hgexpf})--(\ref{hgexpp}) are matched by finding values of $f_0 {r_+}^{2z}$ and $p_0 {r_+}^{2}$. Our numerical calculations are carried out over a range of 4 to 9  dimensional spacetime. The results are shown in Figures~\ref{fig:extfandp1} and ~\ref{fig:extfandp2} where the red dashed line is a plot of the asymptotic expectation and the blue solid line is the numerical result.

\begin{figure}[p]
        \subfloat[Dots are calculated under that $z=2$ and $h_0$ runs from 0.9713 to 0.9705 in increments of $0.00004$ and are joined by straight lines.
         ]{\includegraphics[scale=0.8]{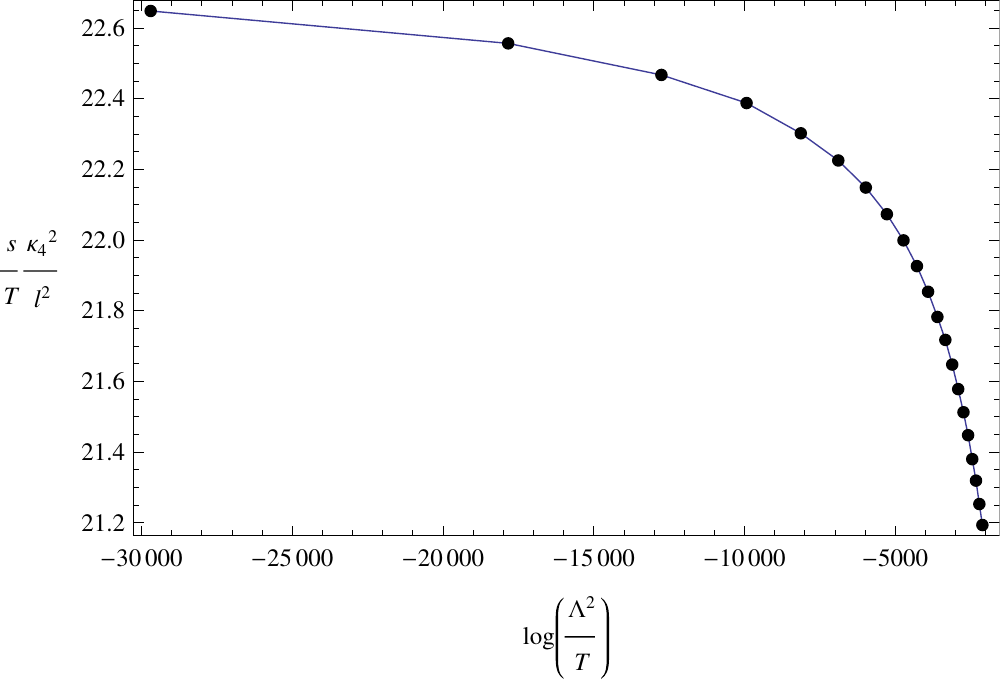}} \; \; \; \; \;
        \subfloat[Dots are calculated under that $z=3$ and $h_0$ runs from 1.6343 to 1.6335 in increments of $0.00004$ and are joined by straight lines.]{\includegraphics[scale=0.8]{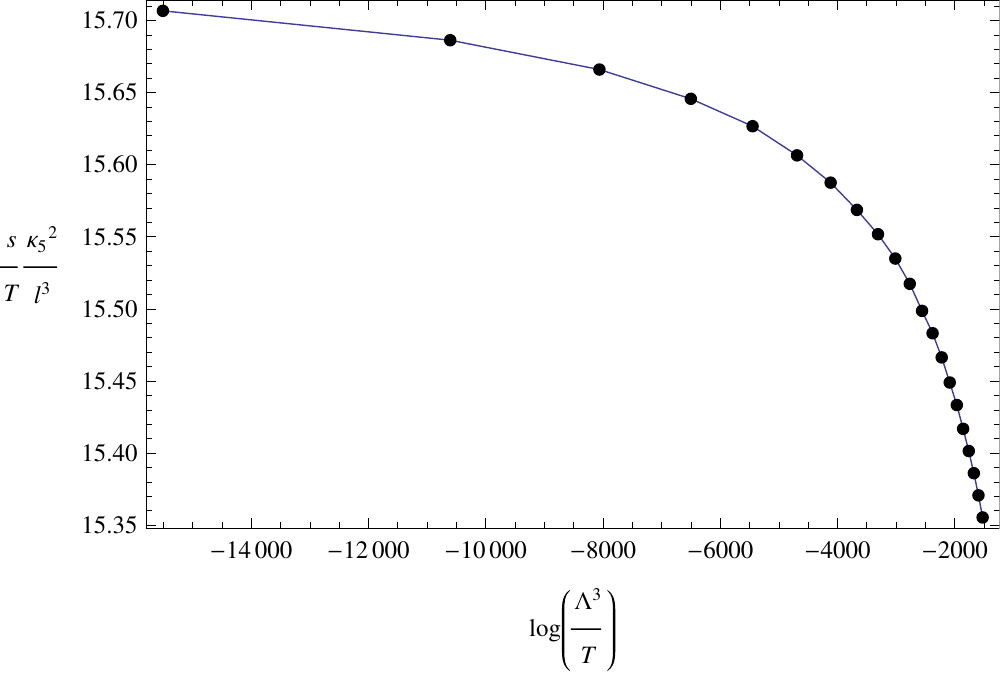}}\\
        \vspace{10pt}\\
        \subfloat[Dots are calculated under that $z=4$ and $h_0$ runs from 2.2822 to 2.2814 in increments of $0.00004$ and are joined by straight lines.]{\includegraphics[scale=0.8]{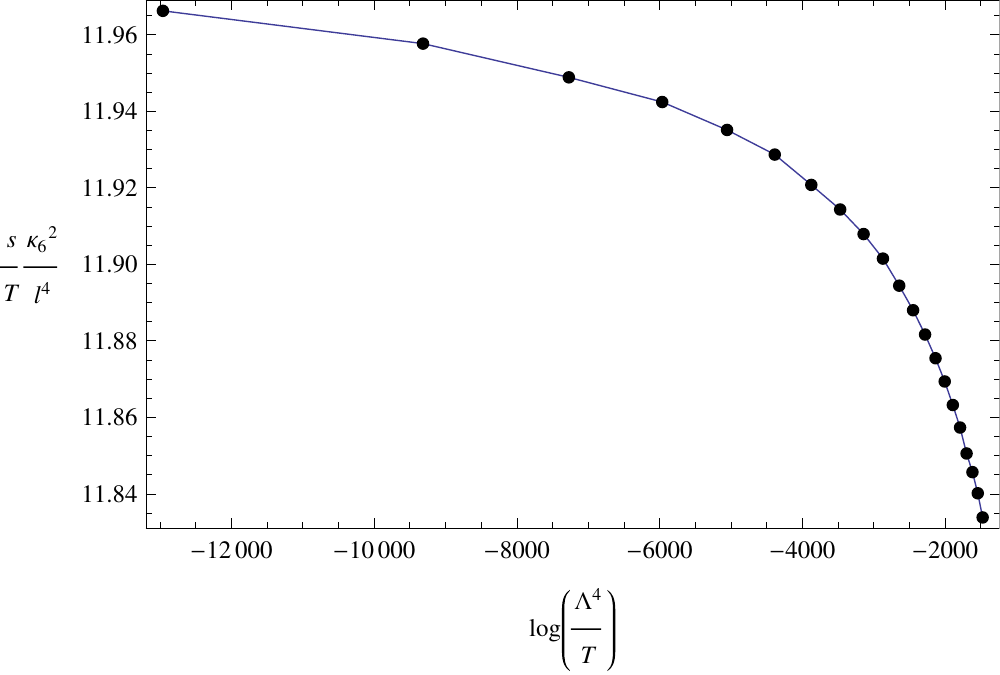}} \; \; \; \; \;
        \subfloat[Dots are calculated under that $z=5$ and $h_0$ runs from 2.9255 to 2.9247 in increments of $0.00004$ and are joined by straight lines.]{\includegraphics[scale=0.8]{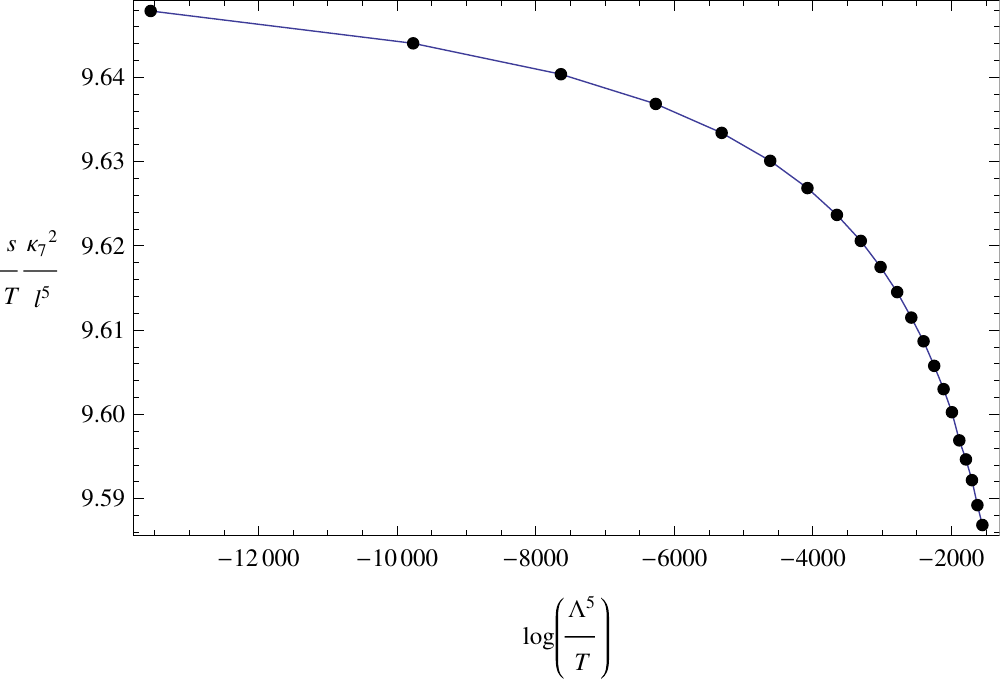}}\\
        \vspace{10pt}\\
        \subfloat[Dots are calculated under that $z=6$ and $h_0$ runs from 3.5668 to 3.566 in increments of $0.00004$ and are joined by straight lines.]{\includegraphics[scale=0.8]{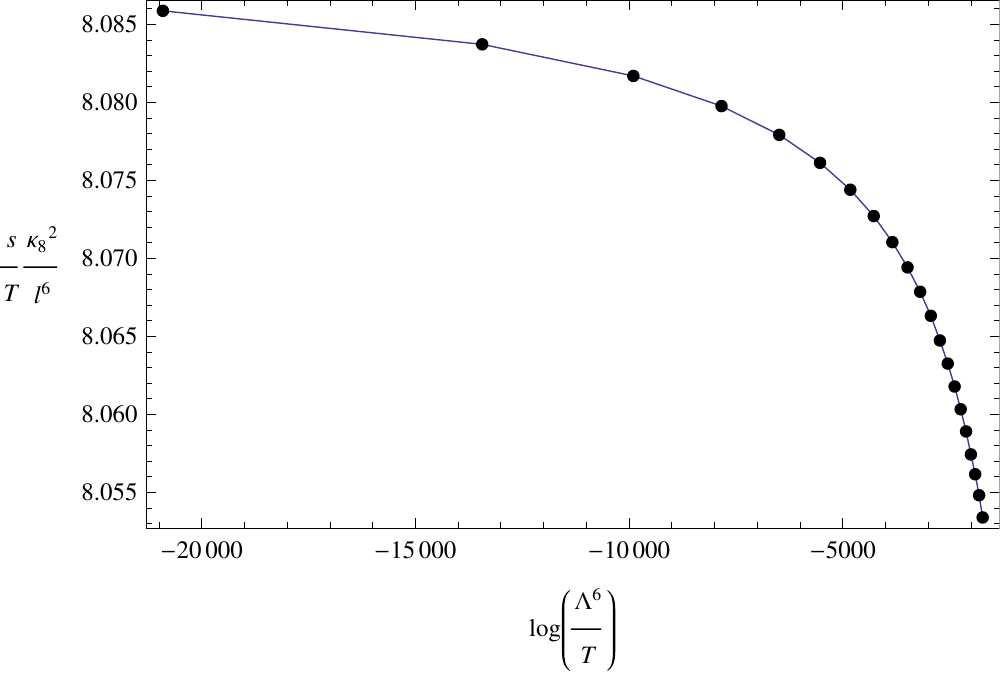}} \; \; \; \; \;
        \subfloat[Dots are calculated under that $z=7$ and $h_0$ runs from 4.2070 to 4.2062 in increments of $0.00004$ and are joined by straight lines.]{\includegraphics[scale=0.8]{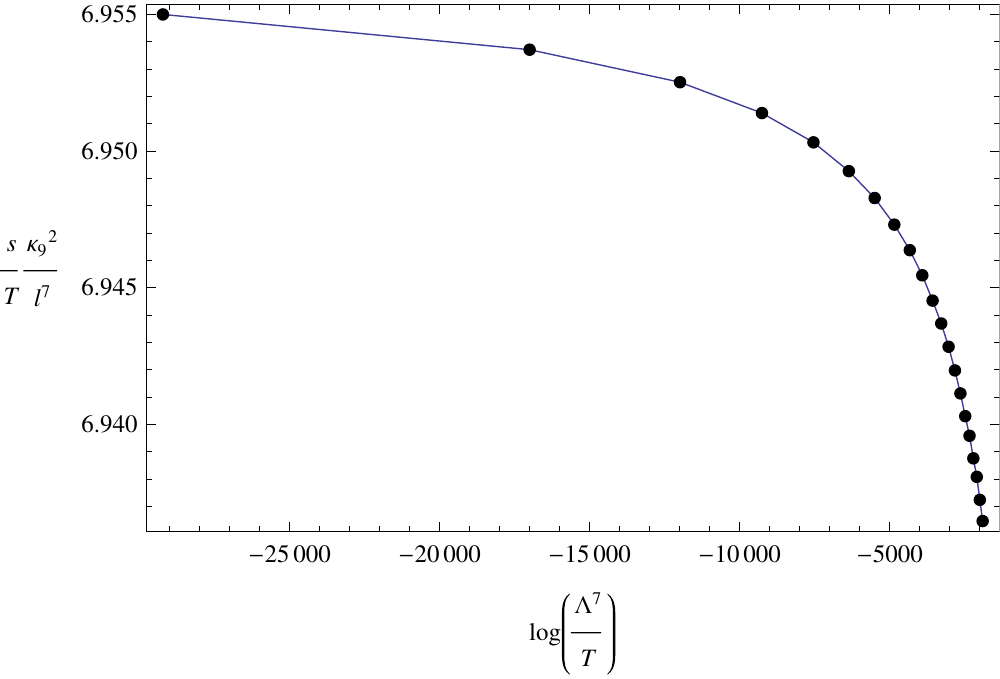}}
        \caption{Entropy density per unit temperature versus $\log(\Lambda^z/T)$}
        \label{fig:entropyD}
\end{figure}

Using the fixed constants $\Lambda$, $f_0$, and $p_0$ we next explore physical quantities of interest. Computing the  entropy density, derived in (\ref{thermoVar}) and determined by $f_0 {r_+}^{2z}$ and $p_0 {r_+}^{2}$, we plot $s/T$ as a function of $\log(\Lambda^z/T)$ in Figure~\ref{fig:entropyD}. Here, the value of $\log(\Lambda^z/T)$, equivalent to  $\log(\Lambda r_+)$, is obtained from the relation between the thermodynamic variables in (\ref{thermoVar}), which is
\begin{equation}
\log \bigg( \frac{\Lambda^z}{T} \bigg) =  z \; \log (\Lambda r_+) + \log(2 \pi) - \frac{1}{2} \log(f_0 {r_+}^{2z}).
\end{equation}
We explicitly consider dimensionalities ranging from 4 to 9.

\subsubsection{Energy Density, and Free Energy Density}
\label{sec:NearHrz}

\begin{figure}[p]
       \subfloat[When $z=2$ and $h_0 = 0.97102$, corresponding to $\log(\Lambda^2/T) \sim -5270$, the left is the numerical result of free energy density ${\mathcal{F}}$ over $Ts$ and the right is the numerical result of energy density ${\mathcal{E}}$ over $Ts$, as a function of $\log(r/r_+)$. The quantity is well defined in the intermediate region. Red dashed line is reading-off the constant value of ${\mathcal{F}}/Ts$ in the intermediate regime.]{\includegraphics[scale=0.8]{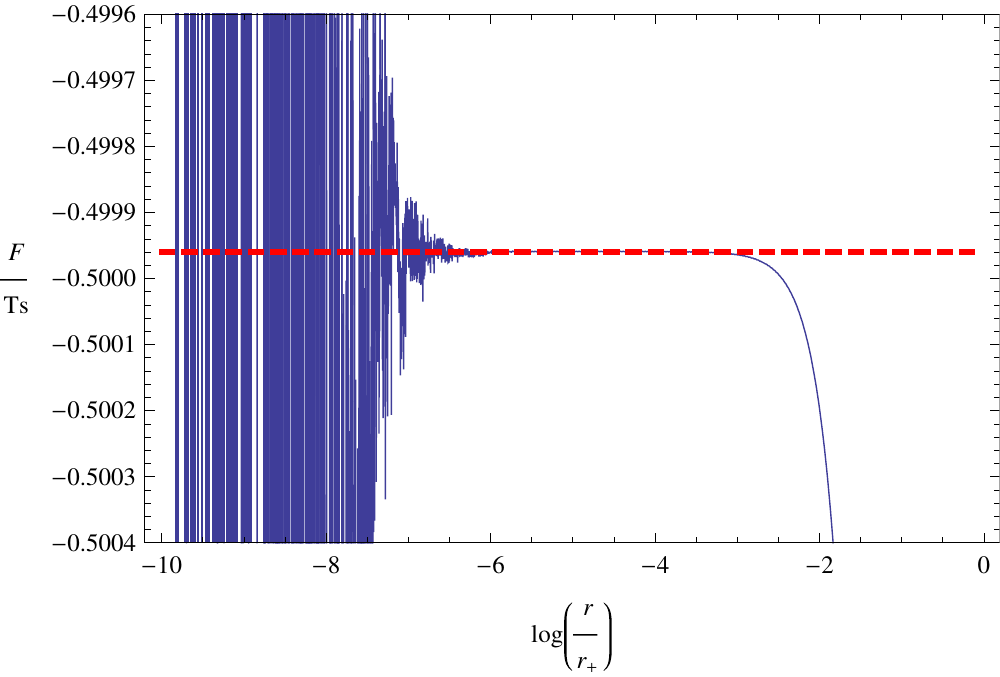}\; \; \; \;\; \includegraphics[scale=0.8]{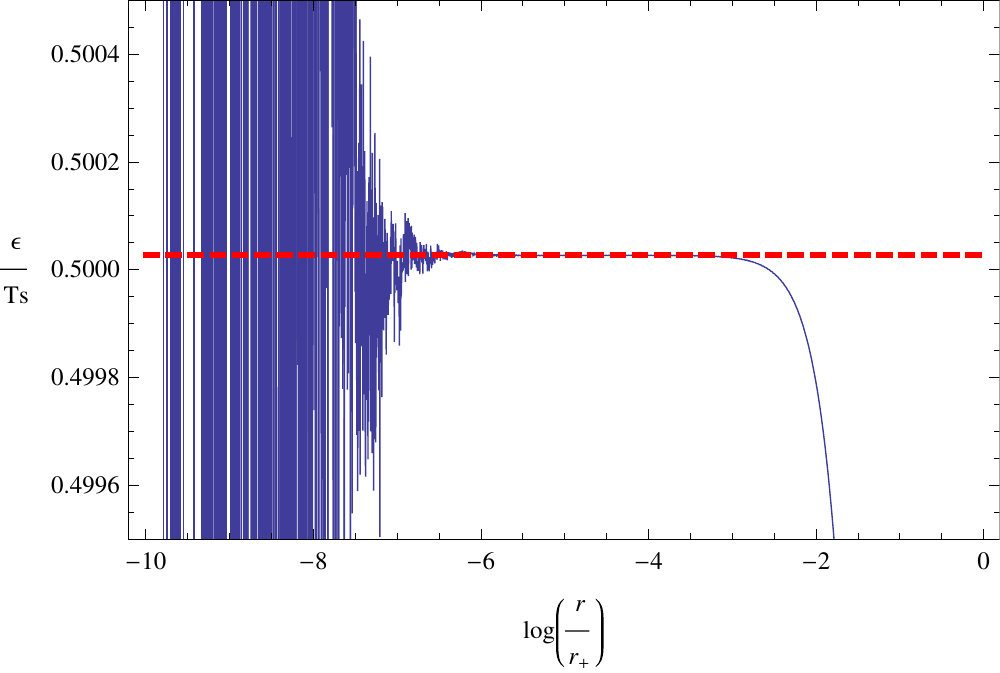}}\\
       \subfloat[When $z=3$ and $h_0 = 1.63422$, corresponding to $\log(\Lambda^2/T) \sim -8000$, the left is the numerical result of free energy density ${\mathcal{F}}$ over $Ts$ and the right is the numerical result of energy density ${\mathcal{E}}$ over $Ts$, as a function of $\log(r/r_+)$. The quantity is well defined in the intermediate region. Red dashed line is reading-off the constant value of ${\mathcal{F}}/Ts$ in the intermediate regime.]{\includegraphics[scale=0.8]{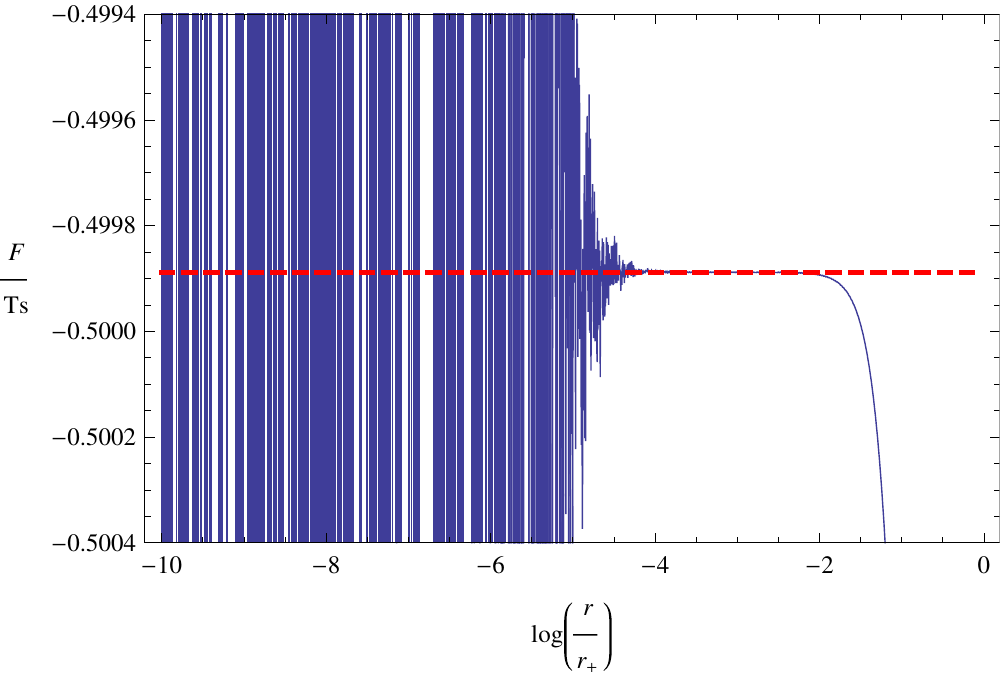}\; \; \; \; \; \includegraphics[scale=0.8]{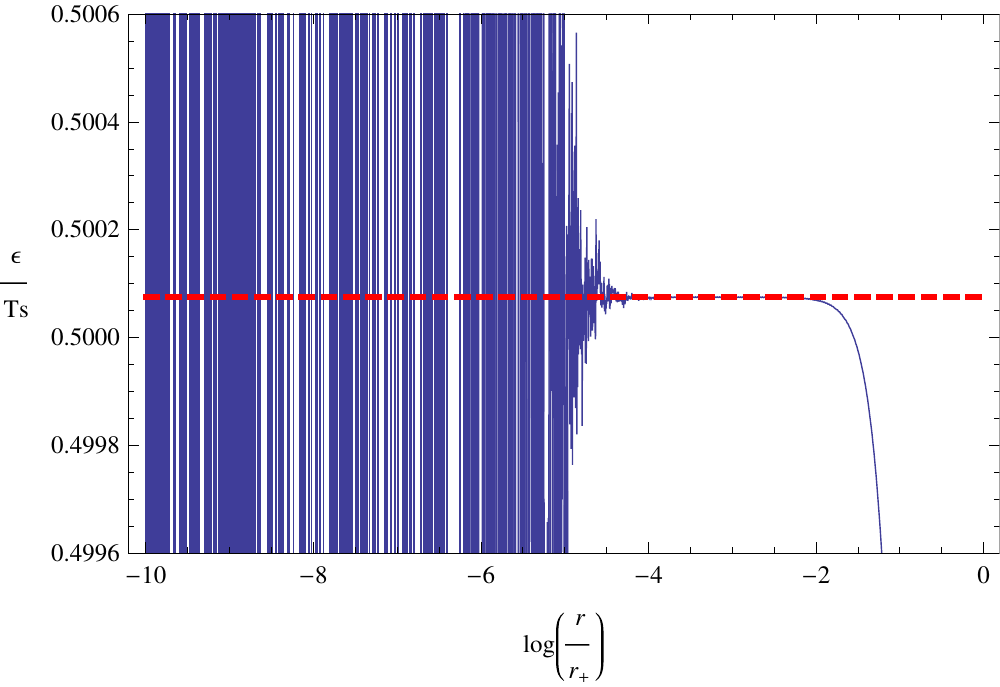}}\\
       \subfloat[When $z=4$ and $h_0 = 2.28212$, corresponding to $\log(\Lambda^2/T) \sim -7000$, the left is the numerical result of free energy density ${\mathcal{F}}$ over $Ts$ and the right is the numerical result of energy density ${\mathcal{E}}$ over $Ts$, as a function of $\log(r/r_+)$. The quantity is well defined in the intermediate region. Red dashed line is reading-off the constant value of ${\mathcal{F}}/Ts$ in the intermediate regime.]{\includegraphics[scale=0.8]{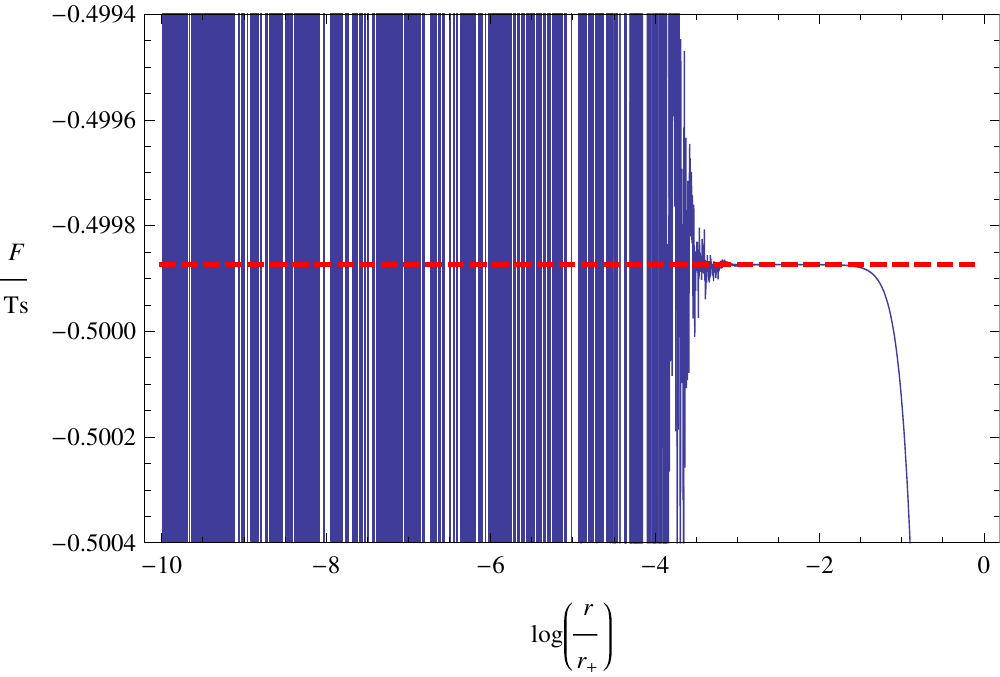}\; \; \; \; \; \includegraphics[scale=0.8]{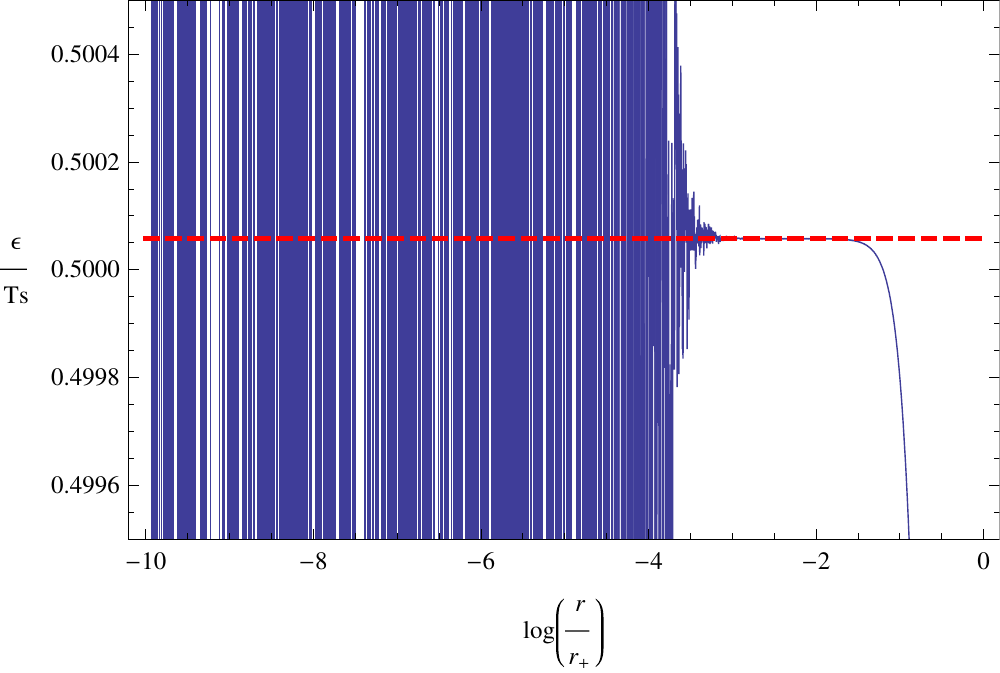}}\\
        \caption{${\mathcal{F}}/Ts$ and ${\mathcal{E}}/Ts$ versus $\log(\frac{r}{r_+})$ for $z=2$, $3$, and $4$}
        \label{fig:densityvsr1}
\end{figure}

\begin{figure}[p]
        \subfloat[When $z=5$ and $h_0 = 2.92526$, corresponding to $\log(\Lambda^2/T) \sim -4100$, the left is the numerical result of free energy density ${\mathcal{F}}$ over $Ts$ and the right is the numerical result of energy density ${\mathcal{E}}$ over $Ts$, as a function of $\log(r/r_+)$. The quantity is well defined in the intermediate region. Red dashed line is reading-off the constant value of ${\mathcal{F}}/Ts$ in the intermediate regime.]{\includegraphics[scale=0.8]{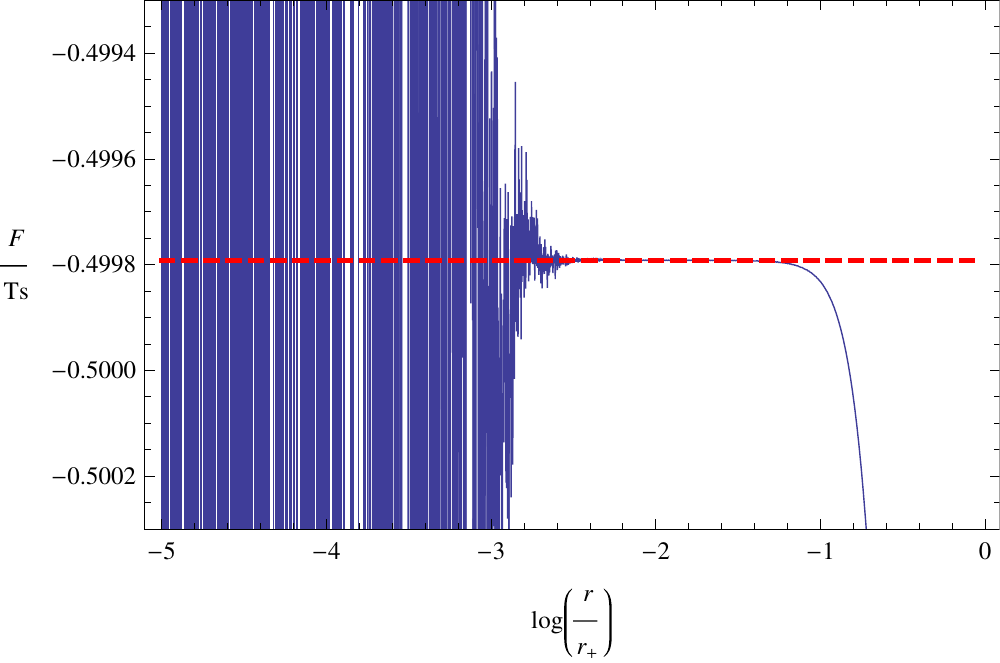}\; \; \; \; \; \includegraphics[scale=0.8]{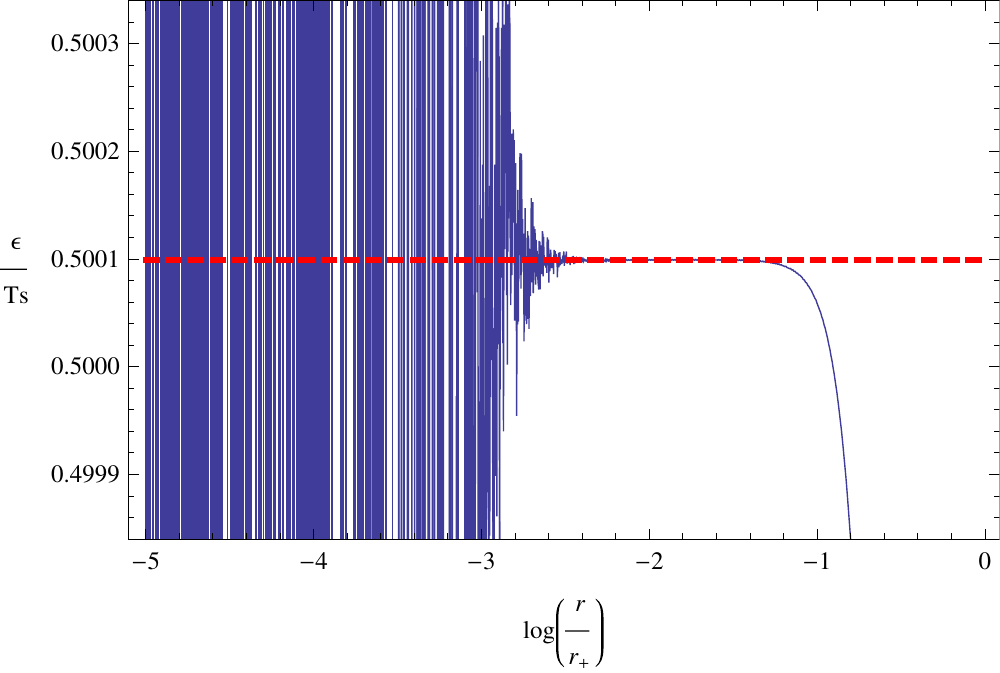}}\\
       \subfloat[When $z=6$ and $h_0 = 3.56670$, corresponding to $\log(\Lambda^2/T) \sim -8700$, the left is the numerical result of free energy density ${\mathcal{F}}$ over $Ts$ and the right is the numerical result of energy density ${\mathcal{E}}$ over $Ts$, as a function of $\log(r/r_+)$. The quantity is well defined in the intermediate region. Red dashed line is reading-off the constant value of ${\mathcal{F}}/Ts$ in the intermediate regime.]{\includegraphics[scale=0.8]{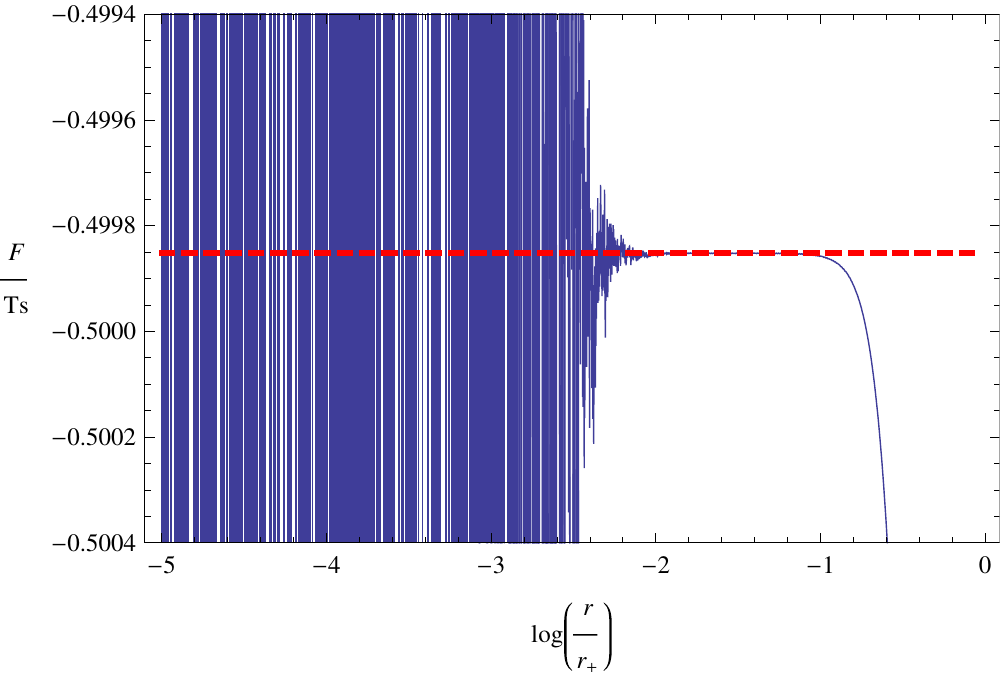}\; \; \; \; \; \includegraphics[scale=0.8]{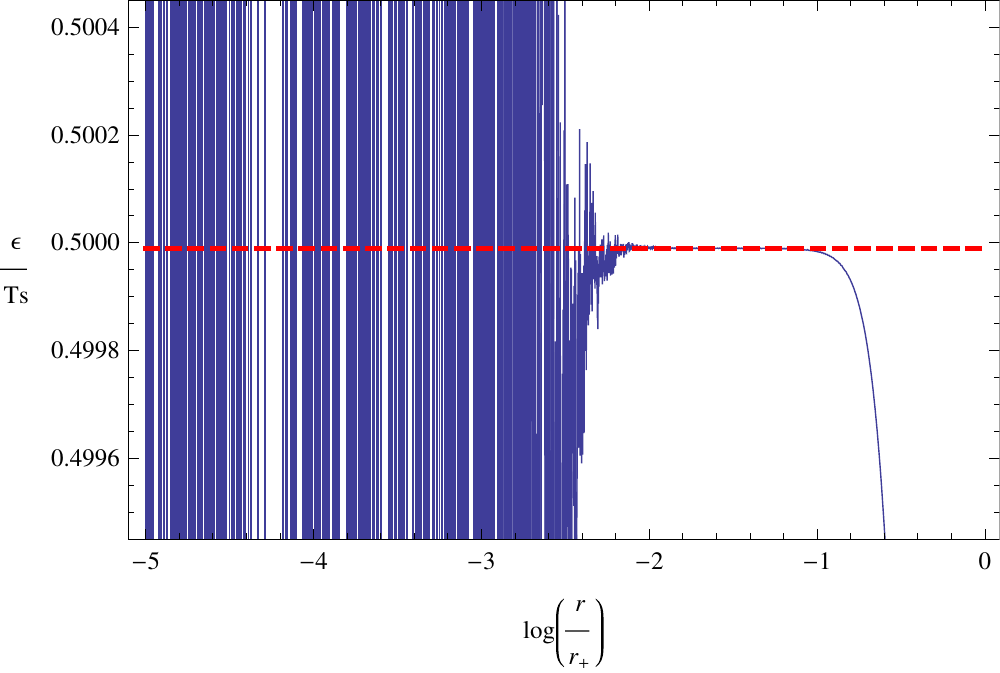}}\\
       \subfloat[When $z=7$ and $h_0 = 4.20694$, corresponding to $\log(\Lambda^2/T) \sim -14000$, the left is the numerical result of free energy density ${\mathcal{F}}$ over $Ts$ and the right is the numerical result of energy density ${\mathcal{E}}$ over $Ts$, as a function of $\log(r/r_+)$. The quantity is well defined in the intermediate region. Red dashed line is reading-off the constant value of ${\mathcal{F}}/Ts$ in the intermediate regime.]{\includegraphics[scale=0.8]{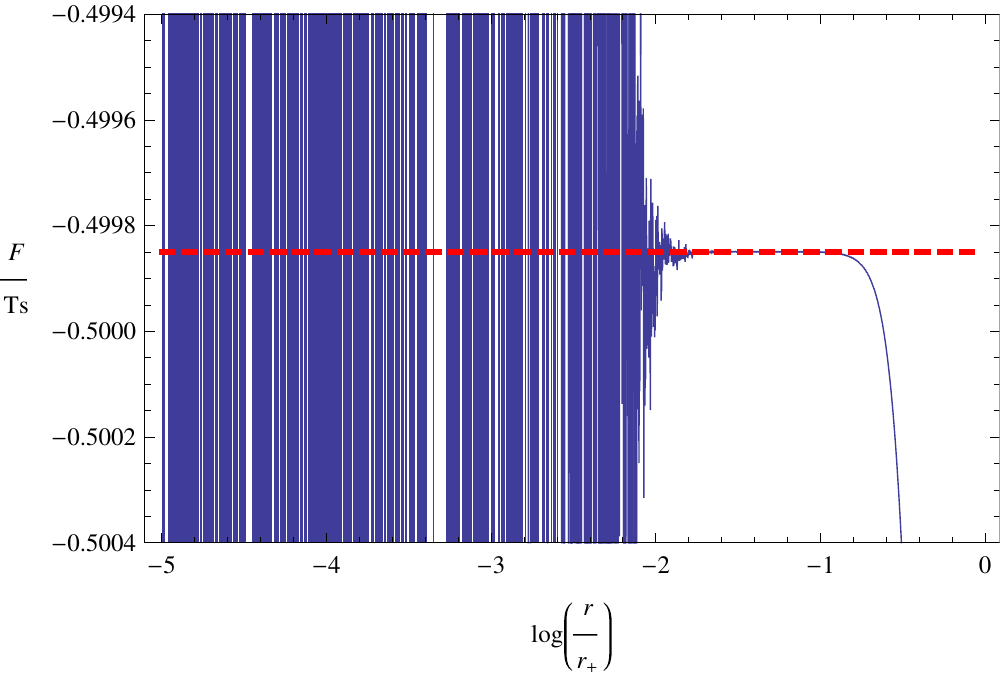}\; \; \; \; \; \includegraphics[scale=0.8]{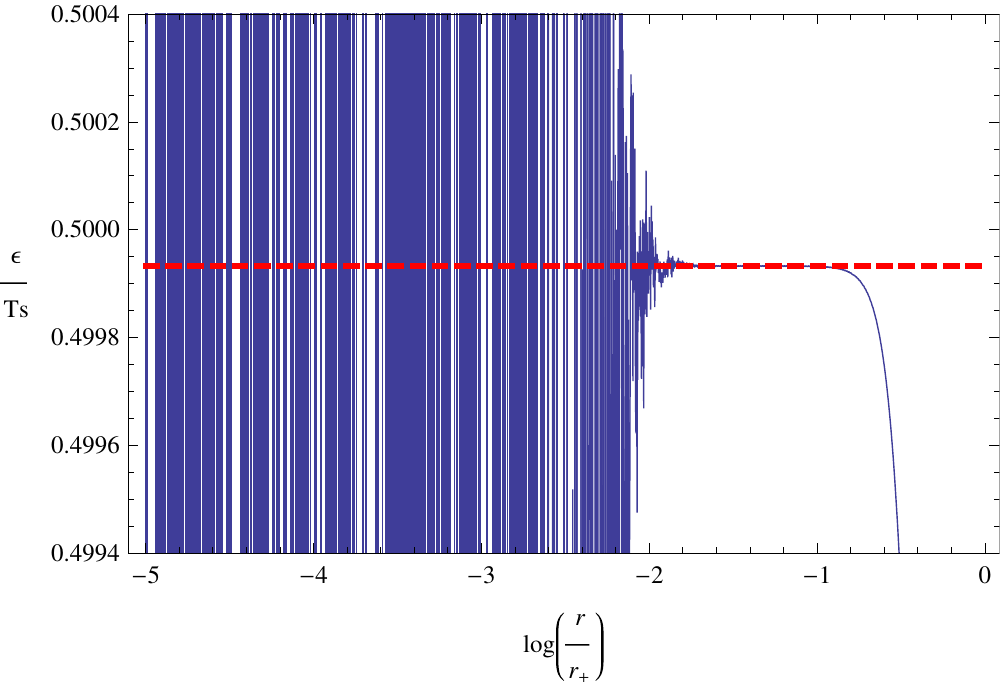}}\\
       \caption{${\mathcal{F}}/Ts$ and ${\mathcal{E}}/Ts$ versus $\log(\frac{r}{r_+})$ for $z=5$, $6$, and $7$}
       \label{fig:densityvsr2}
\end{figure}

Now we move to the free energy density (\ref{defFED}) and energy density (\ref{defED}). Putting the numerical results of $k$, $q$, and $x$ functions into (\ref{defFED}) by using (\ref{newvars}) and (\ref{newx}), and applying the counterterm (\ref{C0}) -- (\ref{C2}), the free energy density and the energy density are respectively depicted  in Figures~\ref{fig:densityvsr1} and~\ref{fig:densityvsr2}. We find that the free energy density and the energy density have a flat region over some finite range of $\log(r/r_+)$, yield a stable constant value for these quantities. In these figures we also illustrate oscillating and divergent behaviours  as the boundary is approached.  Recall that initially the physical quantities having  marginally relevant modes  diverge at the boundary, necessitating the addition of counterterms to render them finite. The  counterterms  should be expanded in an infinite series in $\log(\Lambda r)$. However  in practice  the counterterms (\ref{C0}) -- (\ref{C2}) are truncated at a finite order, while our numerical results include higher orders than ones we considered in the analytic calculation. This limitation is responsible for the unstable behavior near the boundary that appears in Figure~\ref{fig:densityvsr1} and~\ref{fig:densityvsr2}. In other words, the terms in the parenthesis in (\ref{defFED}) and (\ref{defED}) no longer exhibit smooth behavior  as the boundary is approached
but slowly start to oscillate due to the (imperfect) matching between the numerical results and the truncated counterterm power series. These oscillations increase as $r$ decreases, and furthermore   are amplified by the factor of $\sqrt{f} p^{z/2}$, which is rapidly growing when $r \rightarrow 0$. Note that the figures 3 and 4  depict $f r^{2z}/f_0 {r_+}^{2z}$ and $p r^{2}/p_0 {r_+}^2$ and are monotonically decreasing and increasing respectively at small $r$, whereas the functions $f$ and $p$  rapidly grow in that region.

As the spacetime dimension increases we find that the flat region   gets narrower and the divergence behaviour starts more quickly from the horizon. The main reason for this is due to the $\sqrt{f} p^{z/2}$ factor commonly appearing in the free energy density (\ref{defFED}) and the energy density (\ref{defED});  both $\sqrt{f}$ and $p^{z/2}$ vary as  $1/r^{z}$ (as shown in (\ref{hgexpp})) as the boundary $r \rightarrow 0$ is approached, whereas the quantities in the parentheses of (\ref{defFED}) and (\ref{defED}) contain no divergent terms.

\subsection{Exploring the dependence of $\mathcal{E}$ and $\mathcal{F}$ on $\log \Lambda^z/T$}
\label{sec:NearHrz}

\begin{figure}[p]
        \subfloat[Plots of ${\mathcal{F}}/Ts$, ${\mathcal{E}}/Ts$, and ${\mathcal{F}}/{\mathcal{E}}$, as a function of $\log (\Lambda^2/T)$ for $z=2$. Dots are numerical results running $h_0$ from $0.9713$ to $0.9707$, which corresponds to $\log (\Lambda^2/T)$ from about $-30000$ to $-2700$. Solid line is fitting equations in table 1.]{\includegraphics[scale=0.8]{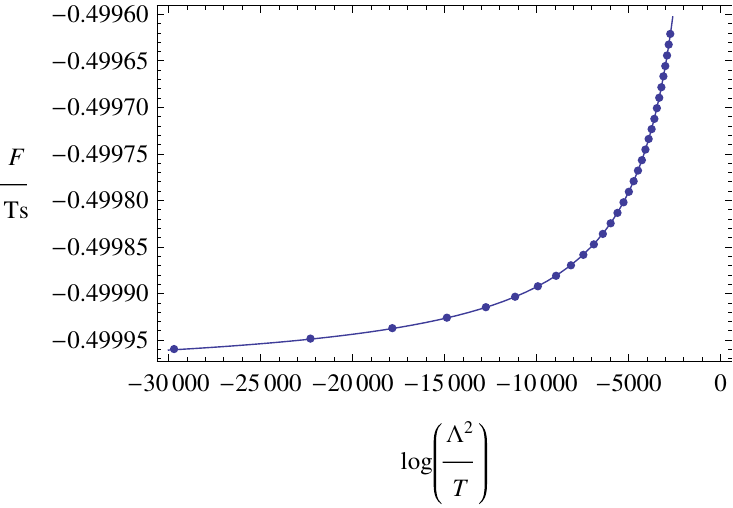} \; \includegraphics[scale=0.8]{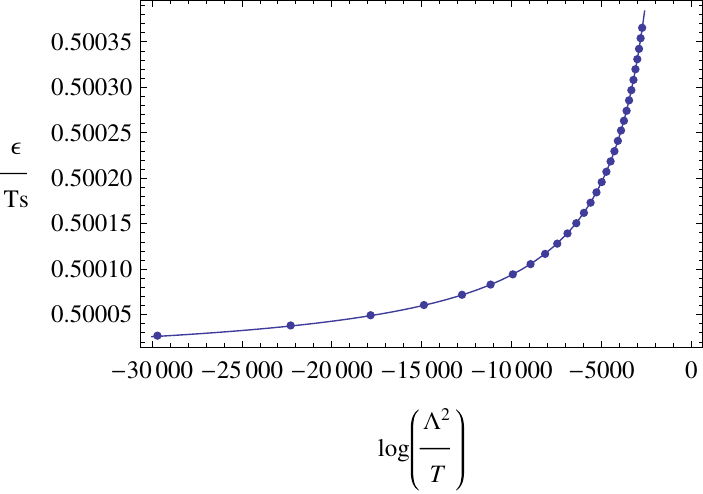} \; \includegraphics[scale=0.8]{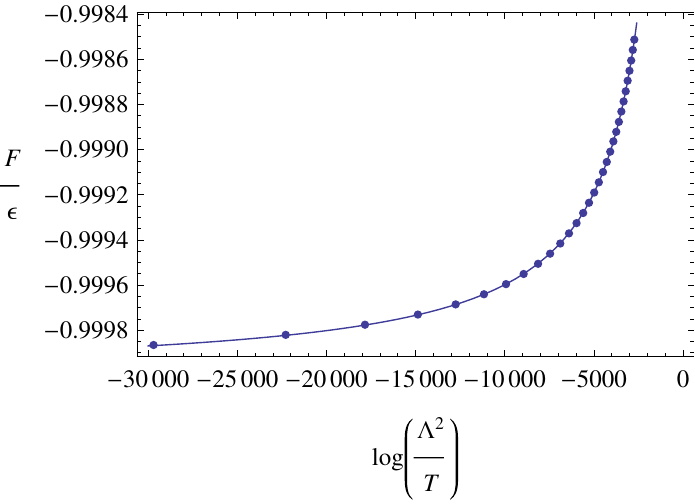}} \\
        \vspace{15pt}\\
        \subfloat[Plots of ${\mathcal{F}}/Ts$, ${\mathcal{E}}/Ts$, and ${\mathcal{F}}/{\mathcal{E}}$, as a function of $\log (\Lambda^3/T)$ for $z=3$. Dots are numerical results running $h_0$ from $1.6343$ to $1.6337$, which corresponds to $\log (\Lambda^3/T)$ from about $-15500$ to $-2000$. Solid line is fitting equations in table 1.]{\includegraphics[scale=0.8]{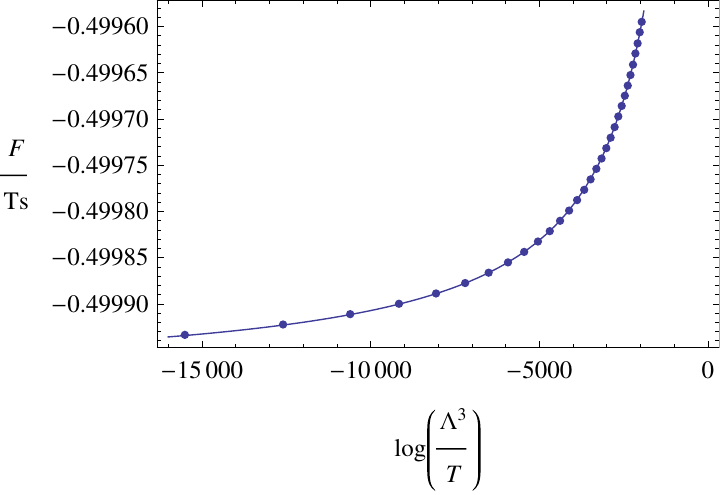} \; \includegraphics[scale=0.8]{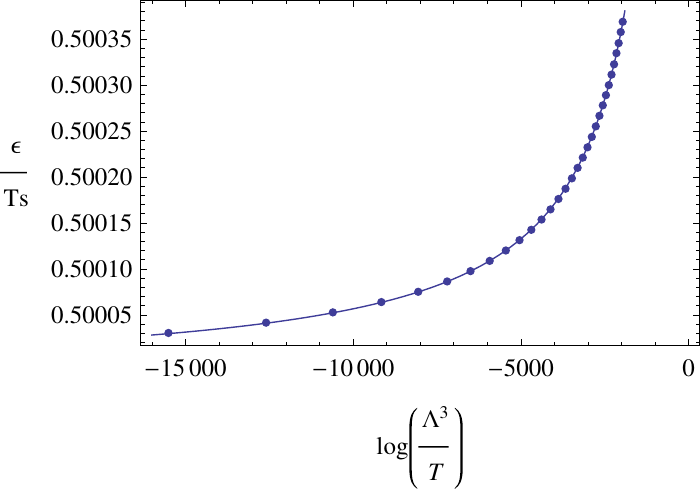} \; \includegraphics[scale=0.8]{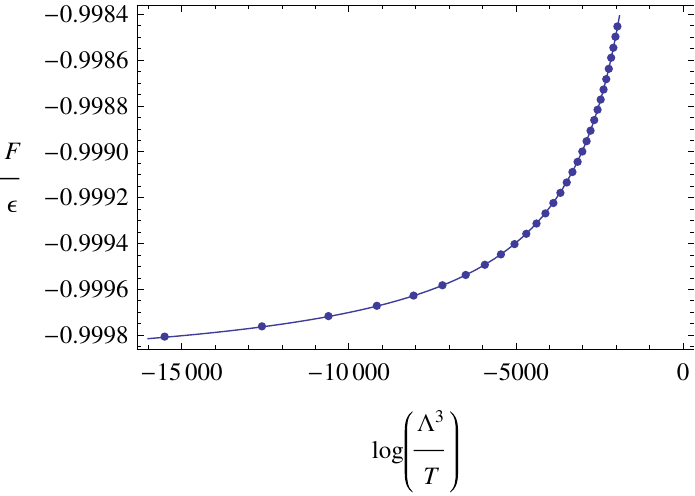}} \\
        \vspace{15pt}\\
        \subfloat[Plots of ${\mathcal{F}}/Ts$, ${\mathcal{E}}/Ts$, and ${\mathcal{F}}/{\mathcal{E}}$, as a function of $\log (\Lambda^4/T)$ for $z=4$. Dots are numerical results running $h_0$ from $2.2822$ to $2.2816$, which corresponds to $\log (\Lambda^4/T)$ from about $-13000$ to $-1900$. Solid line is fitting equations in table 1.]{\includegraphics[scale=0.8]{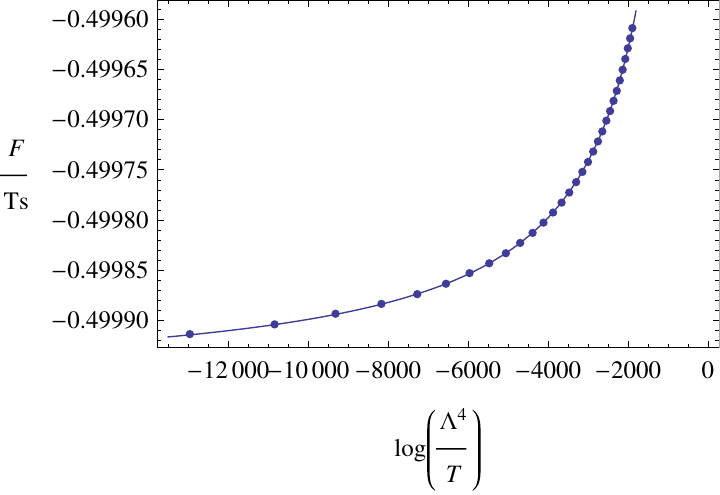} \; \includegraphics[scale=0.8]{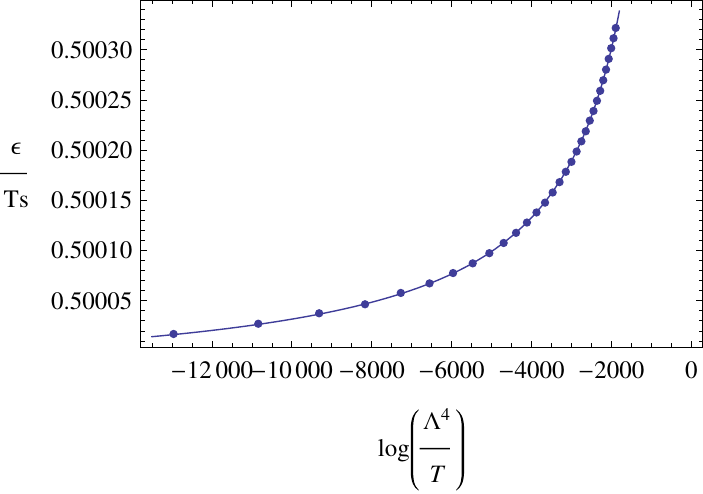} \;  \includegraphics[scale=0.8]{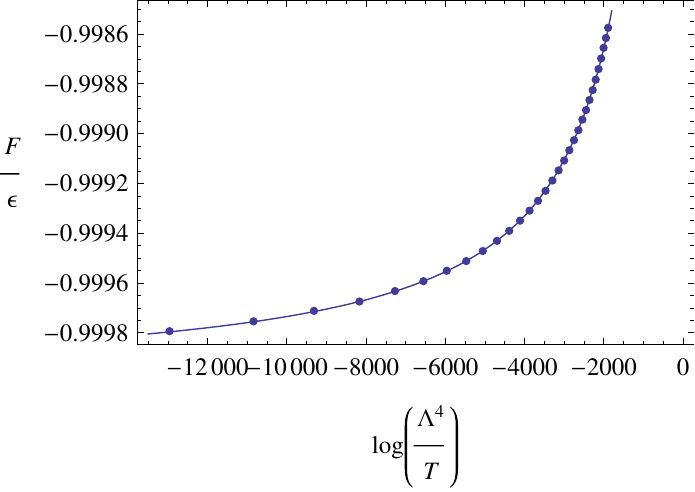}}\\
        \vspace{5pt}\\
        \caption{Plots of ${\mathcal{F}}/Ts$ , ${\mathcal{E}}/Ts$ and ${\mathcal{F}}/{\mathcal{E}}$ versus $\log(\Lambda^z/T)$ for $z=2$, $3$, and $4$.}
        \label{fig:densityvstemp1}
\end{figure}

\begin{figure}[p]
        \subfloat[Plots of ${\mathcal{F}}/Ts$, ${\mathcal{E}}/Ts$, and ${\mathcal{F}}/{\mathcal{E}}$, as a function of $\log (\Lambda^5/T)$ for $z=5$. Dots are numerical results running $h_0$ from $2.9255$ to $2.9249$, which corresponds to $\log (\Lambda^5/T)$ from about $-14000$ to $-2000$. Solid line is fitting equations in table 1.]{\includegraphics[scale=0.8]{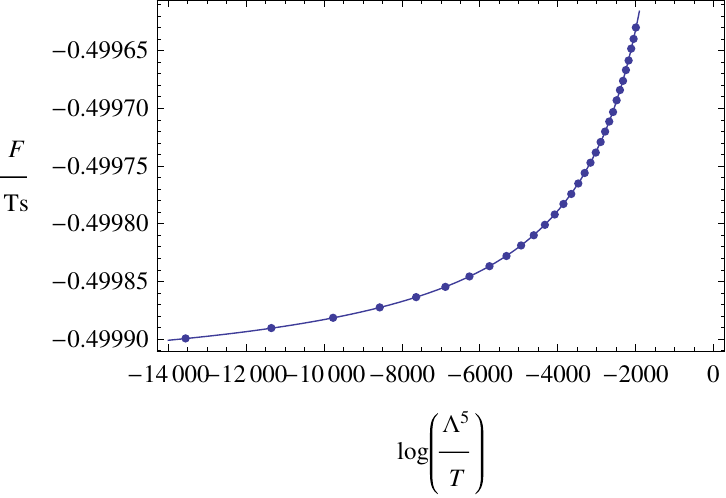} \; \includegraphics[scale=0.8]{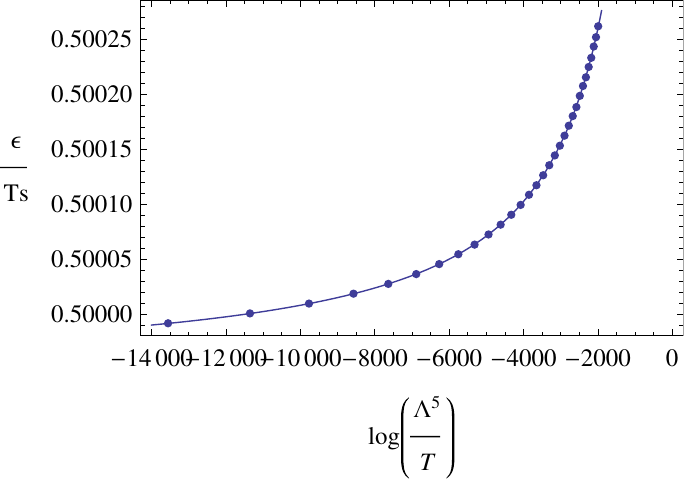} \;\includegraphics[scale=0.8]{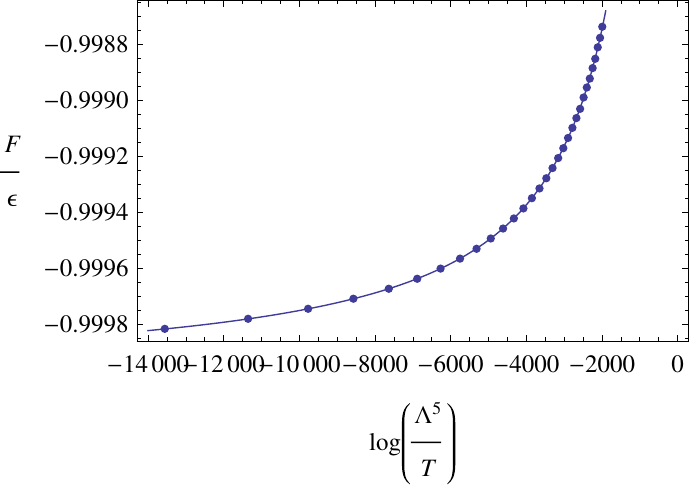}} \\
        \vspace{15pt}\\
        \subfloat[Plots of ${\mathcal{F}}/Ts$, ${\mathcal{E}}/Ts$, and ${\mathcal{F}}/{\mathcal{E}}$, as a function of $\log (\Lambda^6/T)$ for $z=6$. Dots are numerical results running $h_0$ from $3.5668$ to $3.5662$, which corresponds to $\log (\Lambda^6/T)$ from about $-21000$ to $-2200$. Solid line is fitting equations in table 1.]{\includegraphics[scale=0.8]{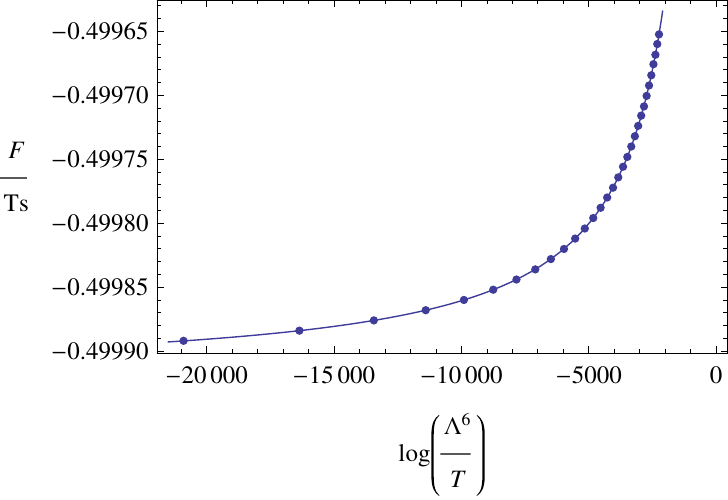} \; \includegraphics[scale=0.8]{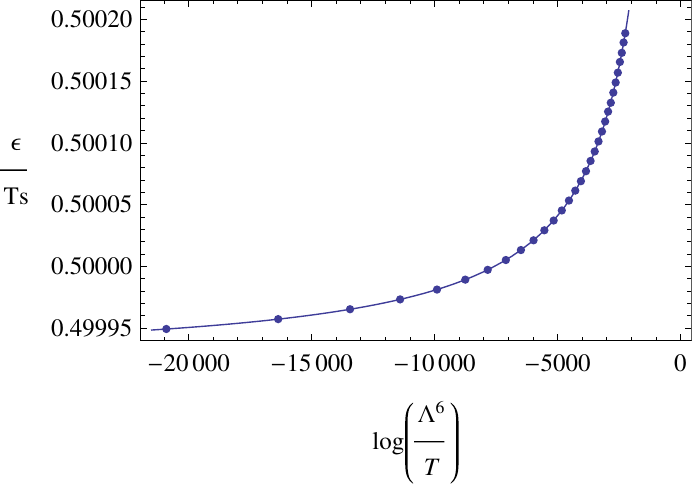} \;\includegraphics[scale=0.8]{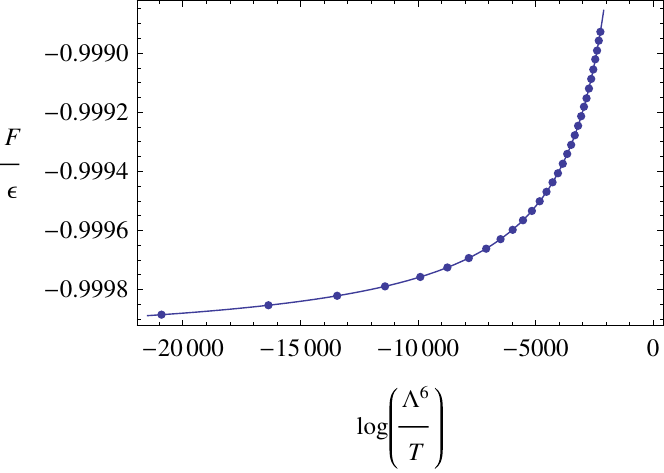}} \\
        \vspace{15pt}\\
        \subfloat[Plots of ${\mathcal{F}}/Ts$, ${\mathcal{E}}/Ts$, and ${\mathcal{F}}/{\mathcal{E}}$, as a function of $\log (\Lambda^7/T)$ for $z=7$. Dots are numerical results running $h_0$ from $4.2070$ to $4.2064$, which corresponds to $\log (\Lambda^7/T)$ from about $-30000$ to $-2200$. Solid line is fitting equations in table 1.]{\includegraphics[scale=0.8]{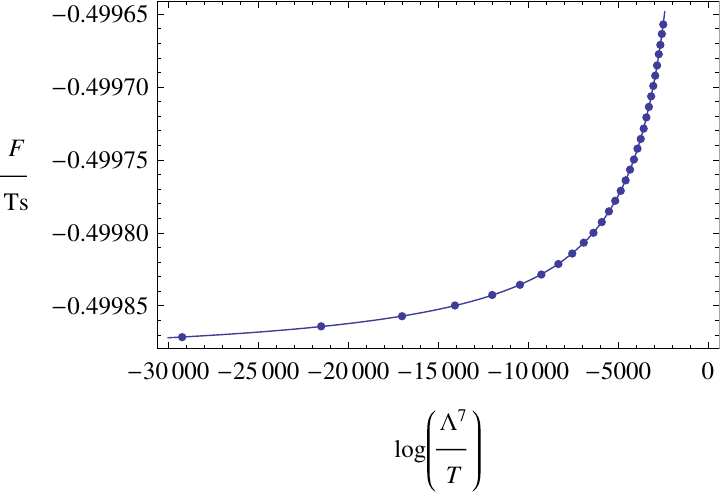} \; \includegraphics[scale=0.8]{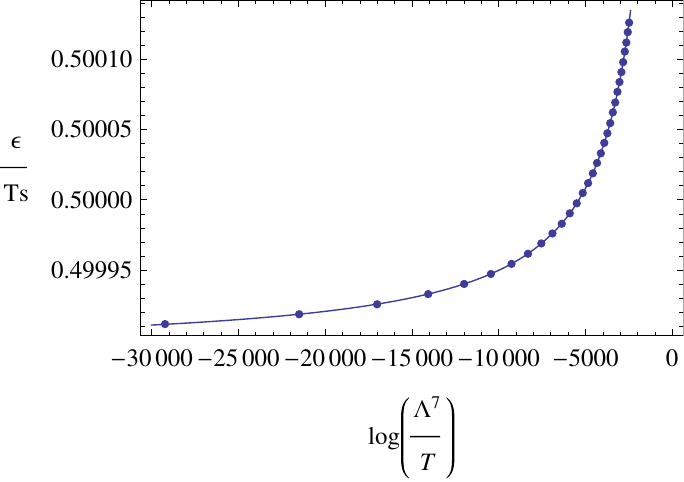} \; \includegraphics[scale=0.8]{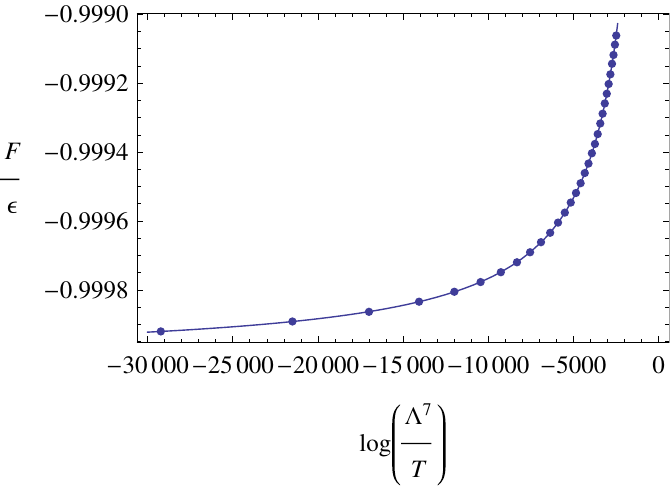}}\\
        \vspace{5pt}\\
        \caption{Plots of ${\mathcal{F}}/Ts$ , ${\mathcal{E}}/Ts$ and ${\mathcal{F}}/{\mathcal{E}}$ versus $\log(\Lambda^z/T)$
        for $z=5$, $6$, and $7$.}
        \label{fig:densityvstemp2}
\end{figure}

In this section, we numerically compute the free energy density per $Ts$ (${\mathcal{F}}/Ts$), and the energy density per $Ts$ (${\mathcal{E}}/Ts$), as functions of $\log(\Lambda^z/T)$, and investigate their behaviour in terms of $T$, assuming that $\Lambda$ is very small and fixed. Figure~\ref{fig:densityvstemp1} and~\ref{fig:densityvstemp2} present our results for these quantities and for   ${\mathcal{F}}/{\mathcal{E}}$.  We also include
the fitting function depicted as a solid line. As expected from analytic considerations for the leading order terms of ${\mathcal{F}}/Ts$ and ${\mathcal{E}}/Ts$ from (5.14) for $\Lambda = 0$, we recover from our numerical results   the same value for the leading order terms. In addition, as the marginally relevant modes generated by $\Lambda \sim 0$ are numerically evaluated, their effect is also shown in Figure~\ref{fig:densityvstemp1} and~\ref{fig:densityvstemp2}. Their behaviour predicts the sub-leading terms  expressed as functions of $\log(\Lambda^z/T)$. For each $z$, the fitting functions of ${\mathcal{F}}/Ts$, $\mathcal{E}/Ts$, and ${\mathcal{F}}/\mathcal{E}$ are presented in Table~\ref{table:fitfunc}.

\begin{figure}[p]
        \subfloat[Plots of $-{\mathcal{F}}/Ts + {\mathcal{E}}/Ts - 1$ for $z=2$ and $h_0$ from $0.9713$ to $0.9707$.]{\includegraphics[scale=0.85]{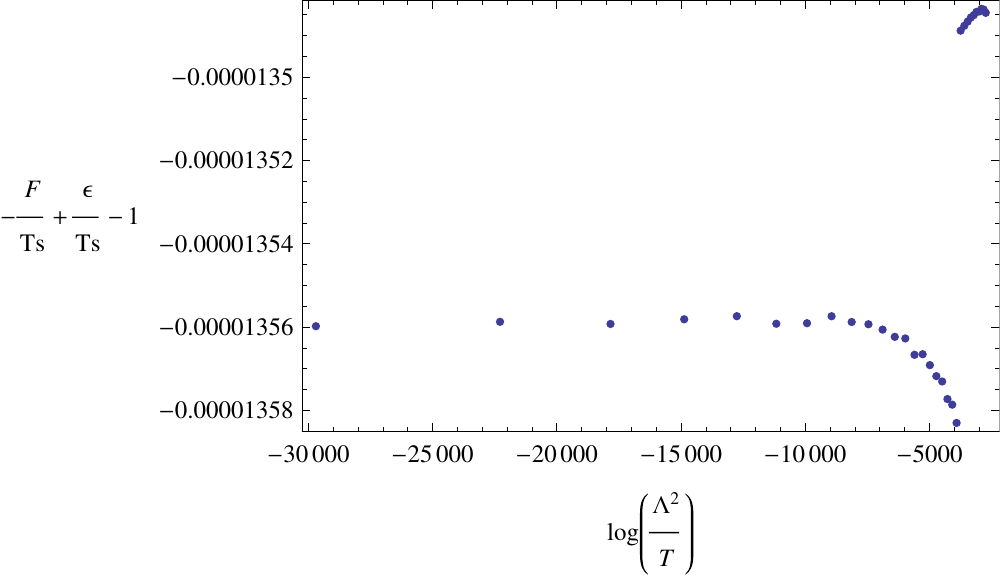}} \; \; \; \; \;
        \subfloat[Plots of $-{\mathcal{F}}/Ts + {\mathcal{E}}/Ts - 1$ for $z=3$ and $h_0$ from $1.6343$ to $1.6337$.]{\includegraphics[scale=0.85]{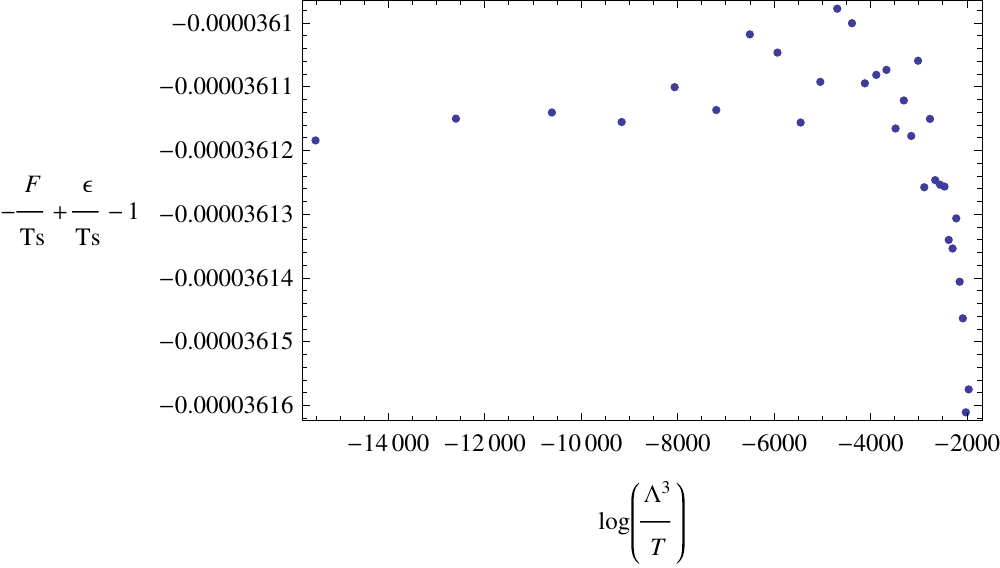}}\\
        \vspace{10pt}\\
        \subfloat[Plots of $-{\mathcal{F}}/Ts + {\mathcal{E}}/Ts - 1$ for $z=4$ and $h_0$ from $2.2822$ to $2.2816$.]{\includegraphics[scale=0.85]{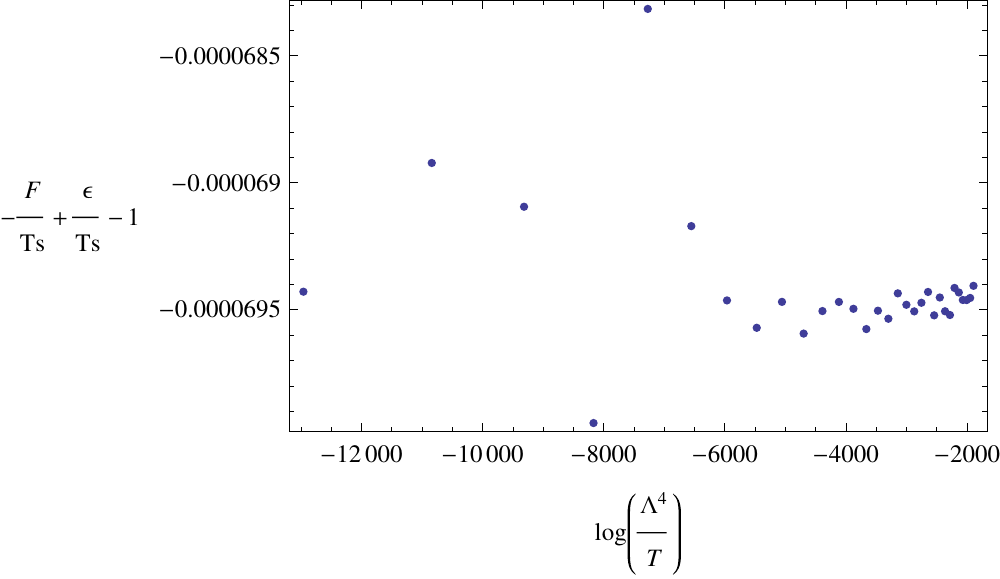}} \; \; \; \; \;
        \subfloat[Plots of $-{\mathcal{F}}/Ts + {\mathcal{E}}/Ts - 1$ for $z=5$ and $h_0$ from $2.9255$ to $2.9249$.]{\includegraphics[scale=0.85]{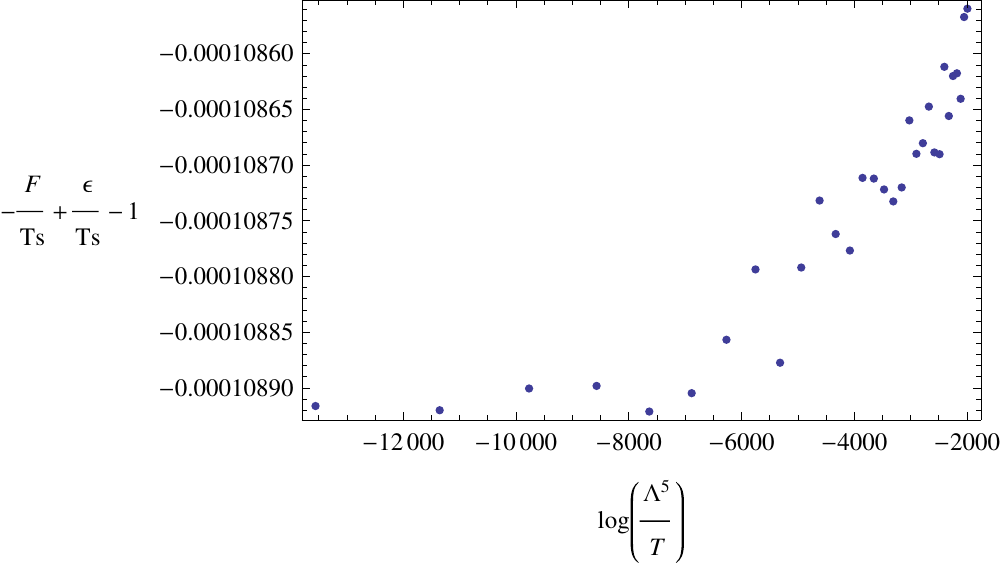}}\\
        \vspace{10pt}\\
        \subfloat[Plots of $-{\mathcal{F}}/Ts + {\mathcal{E}}/Ts - 1$ for $z=6$ and $h_0$ from $3.5668$ to $3.5662$.]{\includegraphics[scale=0.85]{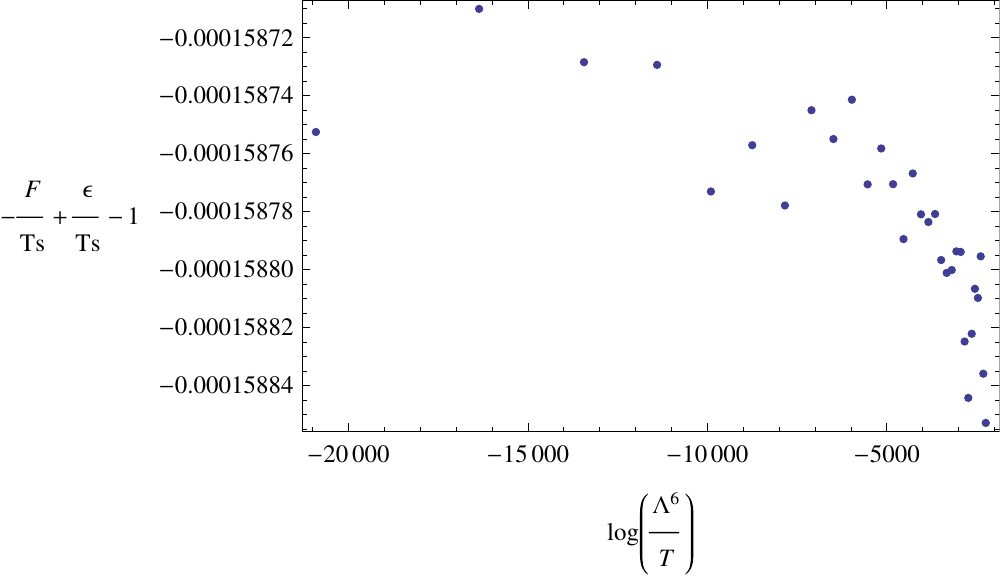}} \; \; \; \; \;
        \subfloat[Plots of $-{\mathcal{F}}/Ts + {\mathcal{E}}/Ts - 1$ for $z=7$ and $h_0$ from $4.2070$ to $4.2064$.]{\includegraphics[scale=0.85]{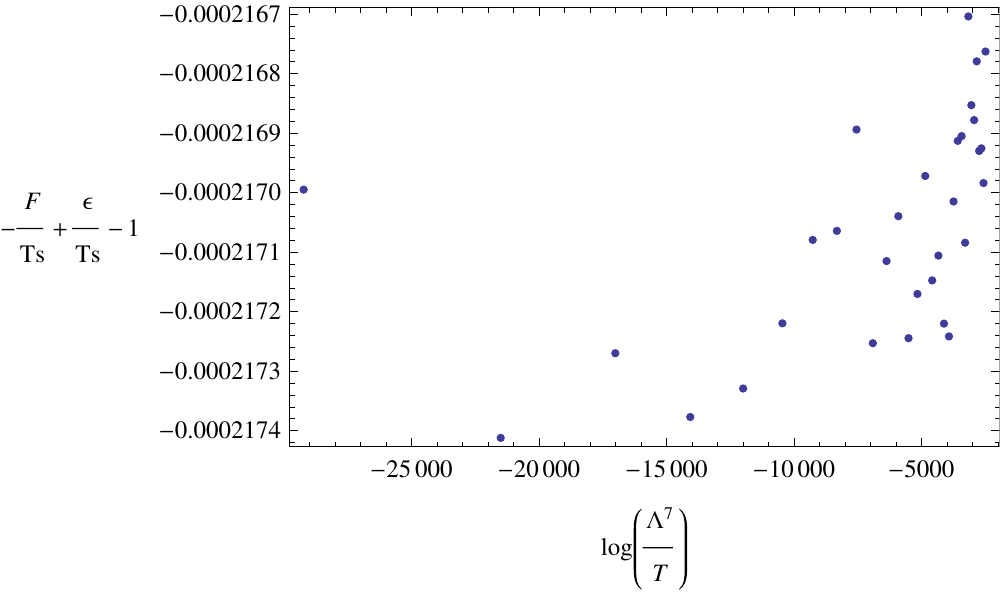}}\\ \vspace{10pt}\\
        \caption{Entropy density over $T$ versus $\log(\Lambda^z/T)$}
        \label{fig:errors}
\end{figure}

In addition, the marginally relevant mode should be consistent with the first law of black hole thermodynamics, which is
$-{\mathcal{F}}/Ts + {\mathcal{E}}/Ts - 1 = 0$. We plot this in Figure~\ref{fig:errors} as a check on the  accuracy on our numerical results. We find that our numerical errors are found between the order of $10^{-3}$ and $10^{-4}$.

\begin{center}
\begin{table}[!h]
  \begin{tabular}{| c || m{4.5cm} | m{4.5cm} | m{4cm} m{0.1cm} |}
    \hline
      & free energy density over $Ts$ $(\frac{{\mathcal{F}}}{Ts}=)$ & energy density over $Ts$ $(\frac{\mathcal{E}}{Ts}=)$ & free energy density over energy density $(\frac{\mathcal{F}}{\mathcal{E}}=)$  & \\[11pt] \hline
    $z=2$ & $ \; \; -\frac{1}{2} - \frac{1}{\log \Lambda^2/T} + \cdots $ & $ \; \; \frac{1}{2} - \frac{1}{\log \Lambda^2/T} + \cdots $ &  $ \; \; -1 - \frac{4}{\log \Lambda^2/T} + \cdots $ & \\[11pt] \hline
    $z=3$ & $ \; \; -\frac{1}{2} - \frac{0.76}{\log \Lambda^3/T} + \cdots $ & $ \; \; \frac{1}{2} - \frac{0.76}{\log \Lambda^3/T} + \cdots $ & $ \; \; -1 - \frac{3}{\log \Lambda^3/T} + \cdots$ & \\[11pt] \hline
    $z=4$ & $ \; \; -\frac{1}{2} - \frac{0.67}{\log \Lambda^4/T} + \cdots $ & $ \; \; \frac{1}{2} - \frac{0.67}{\log \Lambda^4/T} + \cdots $ & $ \; \; -1 - \frac{2.7}{\log \Lambda^4/T} + \cdots$ & \\[11pt] \hline
    $z=5$ & $ \; \; -\frac{1}{2} - \frac{0.63}{\log \Lambda^5/T} + \cdots $ & $ \; \; \frac{1}{2} - \frac{0.63}{\log \Lambda^5/T} + \cdots $ & $ \; \; -1 - \frac{2.50}{\log \Lambda^5/T} + \cdots$ & \\[11pt] \hline
    $z=6$ & $ \; \; -\frac{1}{2} - \frac{0.60}{\log \Lambda^6/T} + \cdots $ & $ \; \; \frac{1}{2} - \frac{0.60}{\log \Lambda^6/T} + \cdots $ & $ \; \; -1 - \frac{2.40}{\log \Lambda^6/T} + \cdots$ & \\[11pt] \hline
    $z=7$ & $ \; \; -\frac{1}{2} - \frac{0.58}{\log \Lambda^7/T} + \cdots $ & $ \; \; \frac{1}{2} - \frac{0.58}{\log \Lambda^7/T} + \cdots $ & $ \; \; -1 - \frac{2.33}{\log \Lambda^7/T} + \cdots$ & \\[11pt] \hline
    \vdots & $\; \; \; \; \; \; \; \; \; \;  \; \; \; \; \; \; \; \; \; \; \vdots$ & $\; \; \; \; \; \; \; \; \; \; \; \; \; \; \; \; \; \; \; \; \vdots $ &  $\; \; \; \; \; \; \; \; \; \; \; \; \; \; \; \; \; \; \; \; \vdots$ & \\
    \hline
    \end{tabular}
    \caption{fitting functions for $\frac{{\mathcal{F}}}{Ts}$, $\frac{\mathcal{E}}{Ts}$, and $\frac{\mathcal{F}}{\mathcal{E}}$}
    \label{table:fitfunc}
\end{table}
\end{center}

\section{Summary and Discussion}
\label{sec:Dscss}

We  have carried out a study of the deformations of Lifshitz spacetime in higher-dimensions, where the deformation is produced by the marginally relevant modes due to $\Lambda \sim 0$, which is the dynamically generated scale. Our main objective was to describe the renormalization group flow of the marginally relevant operators and to investigate the thermodynamic behaviour at finite temperatures
in terms of the dimensionality of the spacetime.   We began with the assumption \cite{Kachru:2008} that the deformation is small in the UV-region and leads to  renormalization group flow into IR-region, where the Lifshitz spacetime is described in the high temperature limit $\Lambda^z/T \rightarrow 0$ and the AdS spacetime is formed in the zero temperature limit $\Lambda^z/T \rightarrow \infty$.

We derived the equation of motion for Einstein gravity with the massive vector field, and set up the ansatz for the metric and the vector potential for both  Lifshitz spacetime and AdS spacetime in section 2. In section 3, we obtained  asymptotic solutions in terms of functions   $k$, $q$, and $x$ and also in functions of $f$, $p$, and $h$ (used separately for later numerical work), and derived the free energy density ${\mathcal{F}}$ and the energy density ${\mathcal{E}}$ at the spacetime boundary in section 4. Since the marginally relevant modes make the action and the physical quantities diverge, we performed holographic renormalization  by adding counterterms, constructed from the covariant quantity $(k^2 - (z-1)/z)$. The coefficients $C_j$ in the counterterm expansion were
obtained as a series in $\log(\Lambda r_+)$.
We also obtained near-horizon expansions of (planar) black hole solutions in section 5.1, and  proved that our analytical results are consistent with the first law of the black hole thermodynamics as $\Lambda = 0$, which means the marginally relevant modes are turned off. Numerically we computed in space-time dimensions 4 to 9 the entropy density $s/T$, the free energy density ${\mathcal{F}}/Ts$, and the energy density ${\mathcal{E}}/Ts$ as functions of  $\log(\Lambda^z/T)$  in near-Lifshitz spacetime.

Our results indicate that the basic physics of Lifshitz/QCT duality \cite{Cheng:2010}  is valid in higher dimensions. Regardless of dimensionality, with a small flux $h_0$, renormalization group flow under the marginally relevant operators for UV-Lifshitz to IR-AdS was obtained in the zero temperature limit $\Lambda^z/T \rightarrow \infty$, commensurate with the 4-dimensional case
{\cite{Kachru:2008}}.  This implies that we can expect   RG flow to exist for each horizon flux, ranging from the zero temperature limit
to  asymptotically Lifshitz black hole spacetimes.  The thermodynamic quantities, $s/T$, ${\mathcal{F}}/Ts$ and ${\mathcal{E}}/Ts$
were calculated as functions of  $\log(\Lambda^z/T)$  just below the maximum value of $h_0$ (which describes a very slightly deformed Lifshitz space-time) and showed graphically similar behaviour.  Also, as analytically expected, the leading-order terms from ${\mathcal{F}}/Ts$ and ${\mathcal{E}}/Ts$ from the first law of black hole thermodynamics for $\Lambda =0$ can be numerically obtained. We also found the sub-leading   terms in terms of  $\log(\Lambda^z/T)$ when $\Lambda \sim 0$. This illustrates
how   physical quantities   change  as functions of $\log(\Lambda^z)/T$ upon approaching
the critical point from a given phase. From a holographic duality perspective  we can characterize the strongly coupled physics having the marginally relevant mode.

Since we considered the case for which the energy-momentum tensor of the gravitational field and the operator of the vector field become marginally relevant, we set $z = n-1$. We consequently found that  higher dimensional Lifshitz black holes can hold more flux, since the maximum value of $h_0$ increased with increasing $n$. The sub-leading term caused by the marginally relevant modes on the ${\mathcal{F}}/Ts$ and ${\mathcal{E}}/Ts$   has a weaker dependence on temperature $T$ as dimensionality increases, as shown in table 2.

There are several possible directions for further research.  One involves
studying the relationship between the RG flows described here and previously noted stability issues in asymptotically Lifshitz
space times  \cite{Copsey:2010ya}.  Another involves consideration of marginally relevant modes in Lifshitz spacetimes with higher curvature corrections. The action with a Gauss-Bonnet term  is now in progress \cite{Park:2012}.

\section*{\bf Acknowledgements}
This work was supported by the Natural Science and Engineering Research Council of Canada.


\appendix



\bibliographystyle{plain}

\begin{thebibliography}{99}

\bibitem{Maldacena:1998}
    Juan Martin Maldacena, \textit{"The Large N limit of superconformal field theories and supergravity."}, Adv.\ Theor.\ Math.\
    Phys.\ {\bf 2}, (1998) 231-252, arXiv:hep-th/9711200.

\bibitem{Son:2008}
    D.T. Son, \textit{"Towward an AdS/cold atoms correspondence: a geometric realization of the Schrodinger symmetry"}, phys.\ Rev.\ D\ {\bf 78:046003}, arXiv:0804.3972 [hep-th].

\bibitem{Gubser:2008}
    S.S. Gubser, and J. McGreevy, \textit{"The gravity dual of a p-wave superconductor"}, JHEP,\ 0811:033,
    (2008), arXiv:0805.2960 [hep-th].

\bibitem{Kachru:2008}
    S. Kachru, X. Liu, and M. Mulligan, \textit{"Gravity duals of lifshitz-like fixed points."}, Phys.\ Rev.\ D\ {\bf 78:106005},
    (2008), arXiv:0808.1725 [hep-th].

\bibitem{Dehghani:2010}
    M. H. Dehghani, and R. B. Mann, \textit{"Lovelock-Lifshitz Black Holes"}, JHEP,\ 1007:019, (2010), arXiv:1004.4397 [hep-th].

\bibitem{Cheng:2010}
    Miranda C.N. Cheng, Sean A. Hartnoll, and Cynthia A. Keeler, \textit{"Deformations of Lifshitz holography."}, JHEP,\ 1003:062,
    (2010), arXiv:0912.2784 [hep-th].


\bibitem{Ross:2009ar}
  S.~F.~Ross and O.~Saremi,
\textit{``Holographic stress tensor for non-relativistic theories''},
  JHEP {\bf 0909}, 009 (2009)
  arXiv:0907.1846 [hep-th].

\bibitem{Balasubramanian:2009}
    K. Balasubramanian and J. McGreevy, \textit{"An analytic lifshitz black hole"}, Phys. \ Rev.\ D\ {\bf
    80:104039}, (2009), arXiv:0909.0263 [hep-th].


\bibitem{Mann:2011hg}
  R.~B.~Mann and R.~McNees,
  \textit{``Holographic Renormalization for Asymptotically Lifshitz Spacetimes''},
  JHEP {\bf 1110}, 129 (2011)
  arXiv:1107.5792 [hep-th].

  \bibitem{Ross:2011gu}
  S.~F.~Ross,
\textit{``Holography for asymptotically locally Lifshitz spacetimes''},
  Class.\ Quant.\ Grav.\  {\bf 28}, 215019 (2011)
  arXiv:1107.4451 [hep-th].

  \bibitem{BravinerGregoryRoss:2011aug}
  H.~Braviner, R.~Gregory and S.~F.~Ross,
  \textit{"Flows involving Lifshitz solutions"},
   Class.\ Quant.\ Grav.\  {\bf 28}, 225028 (2011)  [arXiv:1108.3067 [hep-th]].


\bibitem{Hollands:2005}
    S. Hollands, A. Ishibashi, and D. Marolf, \textit{"Counter-term charges generate bulk symmetries"}, Phys. \ Rev.\ D\ {\bf
    72:104025}, (2005), arXiv:hep-th/0503105

\bibitem{Taylor:2008}
    M. Taylor, \textit{"Non-relativistic holography"}, arXiv:0812.0530 [hep-th]

\bibitem{Bertoldi:2005}
    Gaetano Bertoldi, Benjamin A. Burrington, and Amanda Peet, \textit{"Black Holes in asymptotically Lifshitz spacetimes with arbitrary critical exponent."}, Phys. \ Rev.\ D\ {\bf 80:126003}, (2009), arXiv:0905.3183 [hep-th]

\bibitem{Mann:2009}
    Robert B. Mann, \textit{"Lifshitz Topological Black Holes."}, JHEP,\ 0906:075, (2009) arXiv:0905.1136 [hep-th]

\bibitem{Danielsson:2009}
    U. H. Danielsson and L. Thorlacius, \textit{"Black holes in asymptotically Lifshitz spacetime."}, JHEP,\ 0903:070, (2009) arXiv:0812.5088  [hep-th]

\bibitem{Bertoldi:2009}
    Gaetano Bertoldi, Benjamin A. Burrington, and Amanda W. Peet, \textit{"Thermodynamics of black branes in asymptotically Lifshitz spacetimes."}, Phys. \ Rev.\ D\ {\bf 80:126004}, (2009), arXiv:0907.4755  [hep-th]

\bibitem{Copsey:2010ya}
  K.~Copsey and R.~Mann,\textit{"Pathologies in Asymptotically Lifshitz Spacetimes"}, JHEP {\bf 1103}, 039 (2011), arXiv:1011.3502 [hep-th].

\bibitem{Park:2012}
    Miok Park, and R. B. Mann, \textit{"Generalization of Deformations of Lifshitz holography with Gauss-Bonnet term into $(n+1)$-dimensional spacetime"}, in progress



\end{thebibliography}

\end{document}